\documentclass[preprint,aps]{revtex4}

\usepackage{graphicx}
\usepackage{dcolumn}
\usepackage{bm}
\usepackage{graphics,amsmath,amssymb}
\usepackage{subfigure}
\usepackage{color}
\usepackage{hyperref}
\usepackage{rotating}
\usepackage[normalem]{ulem}
\usepackage{gensymb}
\usepackage [english]{babel}
\usepackage [autostyle, english = american]{csquotes}
\usepackage{array}

\definecolor{maroon}{RGB}{102,0,204}
\definecolor{darkgreen}{RGB}{0,102,51}
\definecolor{orange}{RGB}{255,128,0}
\definecolor{brown}{RGB}{153,76,0}
\definecolor{light-gray}{gray}{0.5}
\definecolor{red}{rgb}{1.0, 0.0, 0.0}

\begin{document}

\preprint{AIP/123-QED}

\title{ Harnessing Leading-Edge Vortices for Improved Thrust Performance of Wave-Induced Flapping Foil Propulsors}
\author{Harshal S. Raut}
 \email{hraut1@jhu.edu}
\affiliation{Department of Mechanical Engineering, Johns Hopkins University, Baltimore, MD, USA}%
\author{Jung-Hee Seo}
 \email{jhseo@jhu.edu}
 \affiliation{Department of Mechanical Engineering, Johns Hopkins University, Baltimore, MD, USA}%
\author{Rajat Mittal}%
 \email{mittal@jhu.edu}
\affiliation{Department of Mechanical Engineering, Johns Hopkins University, Baltimore, MD, USA}%

\date{\today}

\begin{abstract}
This study employs high-fidelity fluid-structure interaction simulations to investigate design optimizations for wave-assisted propulsion (WAP) systems using flapping foils. Building on prior work that identified the leading-edge vortex (LEV) as critical to thrust generation for these flapping foil propulsors, this work explores pitch control mechanisms and foil geometries to improve performance across varying sea states. Two pitch-limiting strategies — a spring-limiter and an angle-limiter — are evaluated. Results show that while both perform similarly at higher sea states, the angle-limiter yields superior thrust at sea-state 1, making it the preferred mechanism due to its simplicity and effectiveness. Additionally, foil geometry effects are analyzed, with thin elliptical and flat plate foils outperforming the baseline NACA0015 shape. The elliptical foil offers marginally better performance and is recommended for WAP applications. A fixed pitch amplitude of 5° provides thrust across all sea states, offering a practical alternative to more complex adaptive systems. These findings demonstrate how insights into the flow physics of flapping foils can inform the design of more efficient WAP systems. 

\end{abstract}

\maketitle

\section{Introduction} \label{sec:intro}
\begin{figure}
    \includegraphics[width=0.8\textwidth]{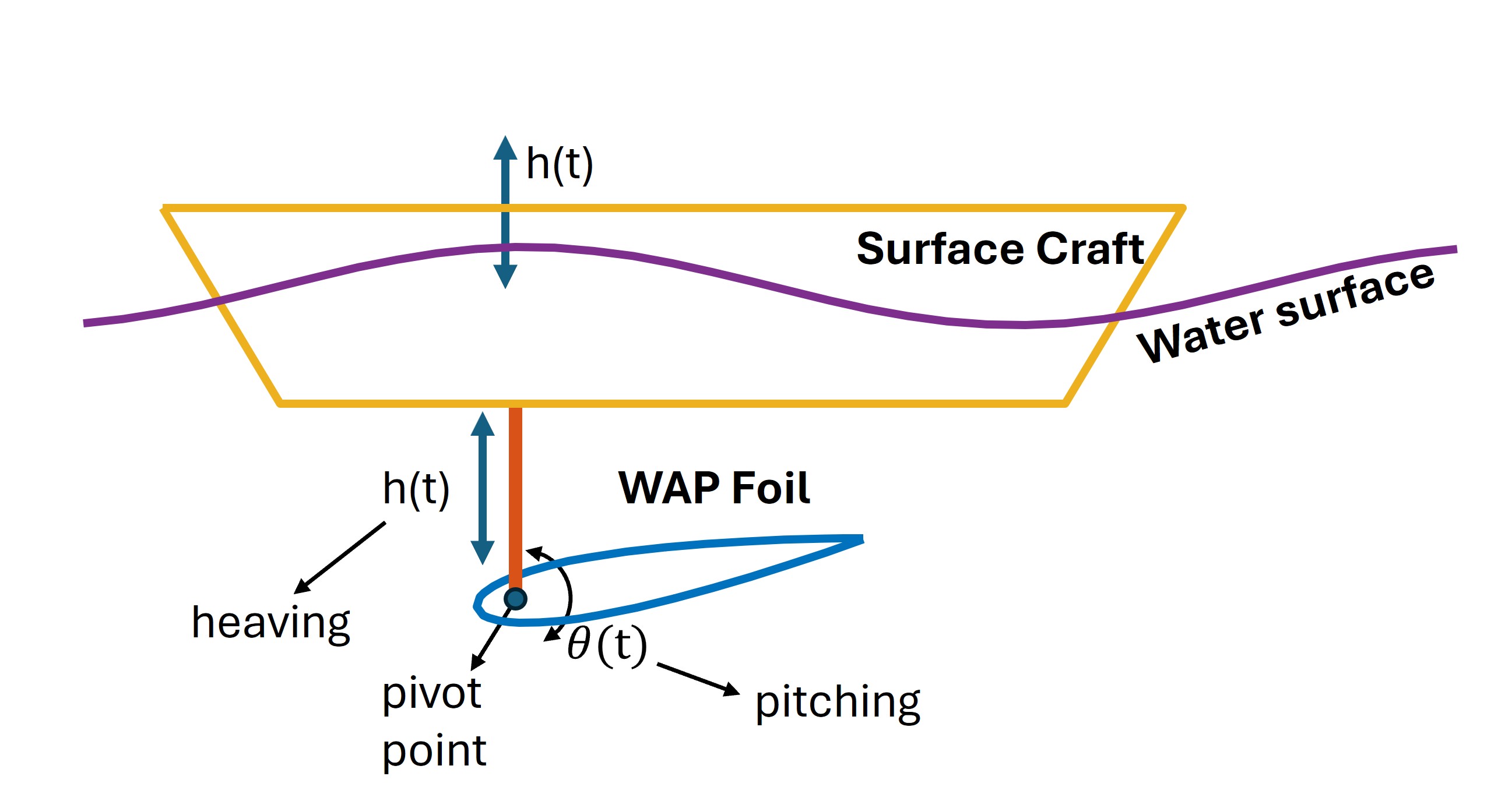}
    \caption{Schematic of surface craft with a wave-assisted propulsion (WAP) system.}
    \label{WAP_Vessel}
\end{figure}

Wave-assisted propulsion (WAP) systems harness the energy of ocean waves to generate propulsion for surface crafts. Various WAP system designs have been investigated \cite{xing2023wave,zhang2024experimental, zhang2022wave, xu2024analysis,qi2020effect,zhang2024dual,yang2019systematic}, most of which include a foil suspended under a surface vehicle (see Fig. \ref{WAP_Vessel}) and attached via a torsional spring to the strut. As ocean waves cause the surface vehicle to oscillate vertically, the submerged foil undergoes periodic heaving motion. The interaction of hydrodynamic forces with the torsional spring's restoring torque induces a ``passive'' cyclic pitching motion in the foil. The synchronized heaving and pitching generate thrust, which can either serve as the primary propulsion for the surface vehicle or enhance its existing propulsion system. 

Several studies have explored passive-pitch WAP systems similar to the one considered here, where wave-induced heave drives the motion and a torsional spring governs passive pitching. Bøckmann and Steen \cite{bockmann2014experiments} conducted experimental investigations, demonstrating that passive-pitch WAP systems can achieve greater efficiency than pitch-controlled flapping hydrofoils.
Using computational fluid dynamics (CFD), Qi et al. \cite{qi2019numerical, qi2020effect} analyzed the influence of spring stiffness, reduced frequency, and mass on the performance of passive-pitch WAP configurations. Yang \emph{et al.} \cite{yang2018numerical, yang2019systematic} examined the hydrodynamic response of flow-induced pitching foils and highlighted the critical role of the torsional spring in the propulsion efficiency. 

In our recent study \cite{raut2024hydrodynamic}, we explored the impact of pivot location and torsional spring stiffness on the thrust performance of WAP systems using energy maps and developed a model based on the phenomenology of the leading-edge vortex (LEV) to predict thrust based solely on foil kinematics. Additionally, we investigated multi-foil configurations\cite{raut2025dynamics}, leveraging the wake of a leading foil to enhance the performance of the trailing foils. We henceforth call this mechanism of using a torsional spring to influence and control passive pitching as a ``spring-limiter'' mechanism. 

The magnitude of thrust generated by a flapping foil is determined by its heaving and pitching kinematics, the frequency of flapping and the forward velocity. In the case of a WAP foil, the heave amplitude is determined by the wave conditions, the forward velocity is connected with the surface vehicle, and the frequency of flapping is equal to the wave encounter frequency, which is determined by the wave conditions and the forward velocity. The pitching kinematics and the shape of the foil used in the WAP propulsor are two design features of the WAP propulsor, and these provide opportunities for devising designs that can maximize thrust over a wide range of operating conditions. Indeed, all the operating parameters can be combined into a single non-dimensional parameter, the wake-width based Strouhal number given by St$_w = f H/U$, where $f$ is the wave encounter frequency, $H$ is the crest-to-trough height of the waves and $U$ is the velocity of the surface craft. As will be shown later, this key parameter can vary over a relatively large range for different sea-states and the design of a WAP propulsor should be such as to enable it to operate effectively over as wide range of this parameter as possible.

\begin{figure}
    \centering
    \includegraphics[width=0.49\textwidth, trim={0.0cm 0cm 0cm 0cm}, clip]{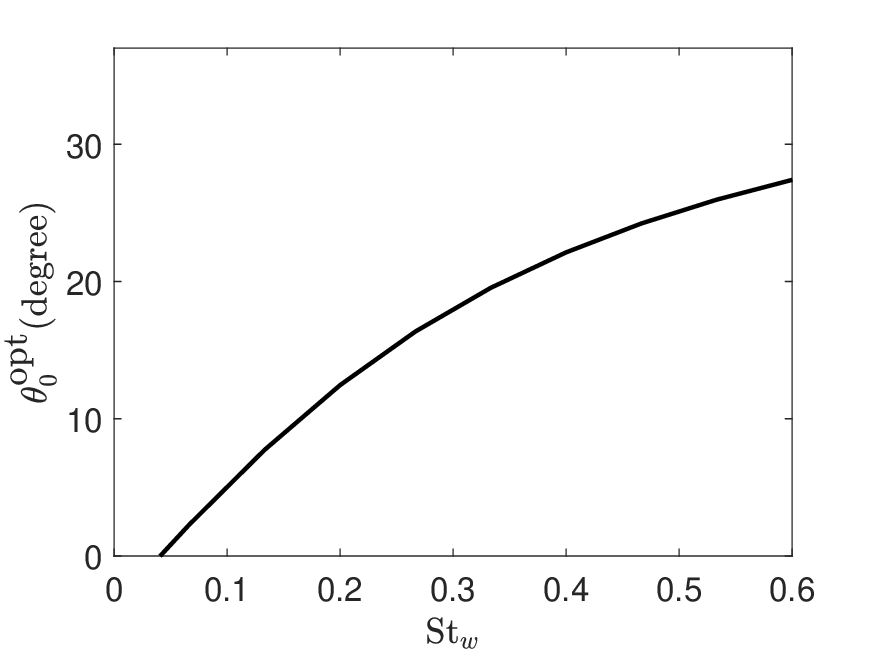}
    \includegraphics[width=0.49\textwidth, trim={0.0cm 0cm 0cm 0cm}, clip]{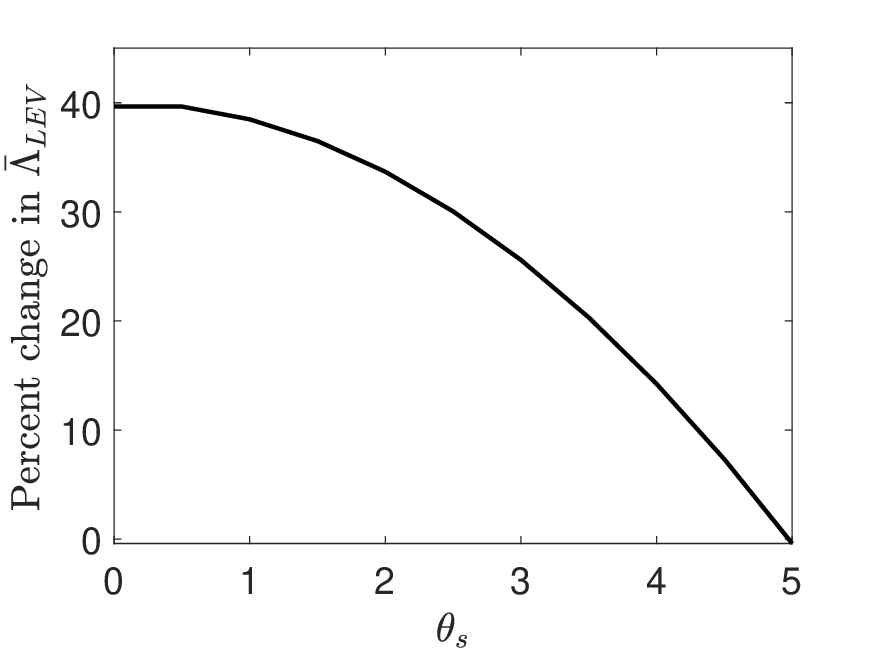}
    \caption{(a) Plot of the optimum pitching amplitude ($\theta_0^\text{opt}$) at various St$_w$ using the LEVBM model for the NACA0015 foil.(b) Plot of percent change in the value of the thrust parameter $\Lambda_{LEV}$ (see Eq. \ref{eq_ctbar} in Appendix A) with $\theta_s$ for St$_w=0.07$ and $H=1$. Thrust is averaged over one cycle and the changes are with respect to $\theta_s=5^\circ$}
    \label{model_prediction}
\end{figure}

In our recent work \cite{raut2024hydrodynamic}, we proposed a LEV (leading-edge vortex)-based model (LEVBM) to predict the thrust generated for heaving and pitching foils. This model was recently used to predict thrust performance in multi-foil\cite{raut2025dynamics} and multi-fish\cite{zhou2025hydrodynamically} configurations as well. The model predicts that for any given Strouhal number, there is a pitch amplitude ($\theta_0$) that maximizes thrust (a detailed derivation of the model is in  Ref \cite{raut2024hydrodynamic} and a succinct description is given in appendix A), which is given by
\begin{equation}
    \theta^\text{opt}_{0} = 0.5 \tan^{-1}\left( \pi \text{St}_w \right)-\theta_s
    \label{theta0_max_eq}
\end{equation}
where the variable $\theta_s$, is an angle-offset which is related to the thickness and shape of the foil at the leading-edge. The value of $\theta_s$ for a NACA0015 foil was estimated from our simulation data to be $\theta_s=3.6^\circ$. 
Fig. \ref{model_prediction} shows the plot of $\theta^\text{opt}_{0}$ versus St$_w$, for a flapping foil to maximize thrust at a given Strouhal number, its pitching motion must be constrained according to the above formula. This prediction has also been verified with high-fidelity flow simulations.

\subsection{Pitch Constraint Mechanisms}
The above suggests that an appropriately designed pitch-constraint mechanism could maximize the performance of a WAP foil. Furthermore, the pitch-constraint mechanism that allows for the maximum pitch-angle to be varied depending on the sea-state conditions (which can be quantified in terms of the Strouhal number) could maximize thrust over a wide range of sea-states. A common pitch-constraint mechanism in WAP propulsors is a torsional spring \cite{xing2023wave,zhang2024experimental,zhang2024dual, yang2019systematic}. However, it is relatively difficult to adjust the stiffness of torsional springs and the relationship between the spring stiffness and the maximum pitch angle is relatively complex. The spring can also lead to complex oscillatory behavior which might diminish thrust. These factors, in addition to the practical aspects of the immersion of the spring mechanism into sea water make the application of torsional springs a less than ideal solution for imposing the pitch-constraint. 

A simpler design is an adjustable pitch-angle limiter (or ``angle-limiter'') that directly imposes a mechanical limit on the pitch-angle without springs. With such an angle-limiter, the foil is allowed to pitch freely under the influence of hydrodynamic forces but is restricted from exceeding a certain pitching angle in either direction. The design of an adjustable angle-limiter would be simple and could also be robust to seawater exposure. However, to the best of our knowledge, no study has yet investigated the use of an angle-limiter as a mechanism to maximize thrust by constraining the pitching motion of a WAP foil. In particular, a WAP foil with an angle-limiter is expected to exhibit kinematics at any given operational condition that are different from those of a foil with a torsional spring, and the effect of this on the fluid-structure interaction and thrust performance of the foils is unclear.

\subsection{Foil Shape Effects}
Eq. \ref{theta0_max_eq} also indicates that the angle $\theta_s$ plays an important role in determining the thrust performance of the flapping foils. This parameter is associated with the shape of the leading-edge of the foil. For the NACA0015 foil that was used in our previous study, $\theta_s$ was estimated from our simulation data to be $3.6^\circ$. In Fig.\ref{model_prediction} (b), we plot percent change in the value of $\bar{\Lambda}_\text{LEV}$ with $\theta_s$ at the optimal value of pitch angle amplitude ($\theta_0^\text{opt}$) where $\bar{\Lambda}_{LEV}$ is a factor directly proportional to the mean thrust coefficient (see Eq. \ref{eq_ctbar} in Appendix A). Thus, foil shapes that correspond to lower values of $\theta_s$ should lead to enhancement in thrust from the foil. Some past studies have examined the effect of shape of the foil on thrust generation. Butt \emph{et al.}\cite{butt2024effect} studied the effect of airfoil camber and reflex camber on flapping foil performance in both energy extraction and propulsion regimes. They found that airfoils with maximum camber location near the leading-edge of the airfoil have higher average lift coefficient and average drag coefficient. Ashraf \emph{et al.}\cite{ashraf2011reynolds} studied the effect of airfoil thickness for pure plunging and combined pitching and plunging foils. They found that for higher Re, significant gains could be achieved both in thrust generation by using a thicker airfoil section for plunging and combined motion with low pitch amplitude. The above studies have however mostly tried to find a correlation between airfoil shape and aerodynamic forces without a direct connection with the flow physics associated with such correlations. To the best of our knowledge the effect of foil shape, especially the shape of the leading-edge, on the LEV-associated thrust generation, has not been investigated before.

\subsection{Objectives}
Motivated by the importance of the LEV for thrust generation of flapping foils, this study employs numerical simulations to examine two design features of WAP propulsors. First we examine the performance of the two pitch-constraint mechanisms - a torsional spring-limiter and an angle-limiter - on thrust performance for a range of sea-states. We perform high-fidelity CFD simulations to study the effect of stiffness of the torsional spring and the maximum angle allowed by the angle-limiter for a range of sea-states that are characterized by different wave heights and frequencies. 
We subsequently examine foils of different shapes, with particular focus on the shape of the leading-edge, and compare their thrust performance for a range of sea-states. Overall, a total of 144 simulations (48 for the pitch-constraint mechanism and 96 for the study of the effect of the foil-shape) have been performed in this study, and these provide a detailed quantitative assessment of these design features and also enable us to examine the underlying hypothesis regarding the LEV-driven thrust performance of flapping foils.

\section{\label{setup}Methodology}
\subsection{\label{prob_setup}Problem Set-up}
\begin{figure}
    \centering
    \subfigure[]
    {\includegraphics[width=0.35\textwidth,trim={4cm 1cm 9.7cm 3cm},clip]{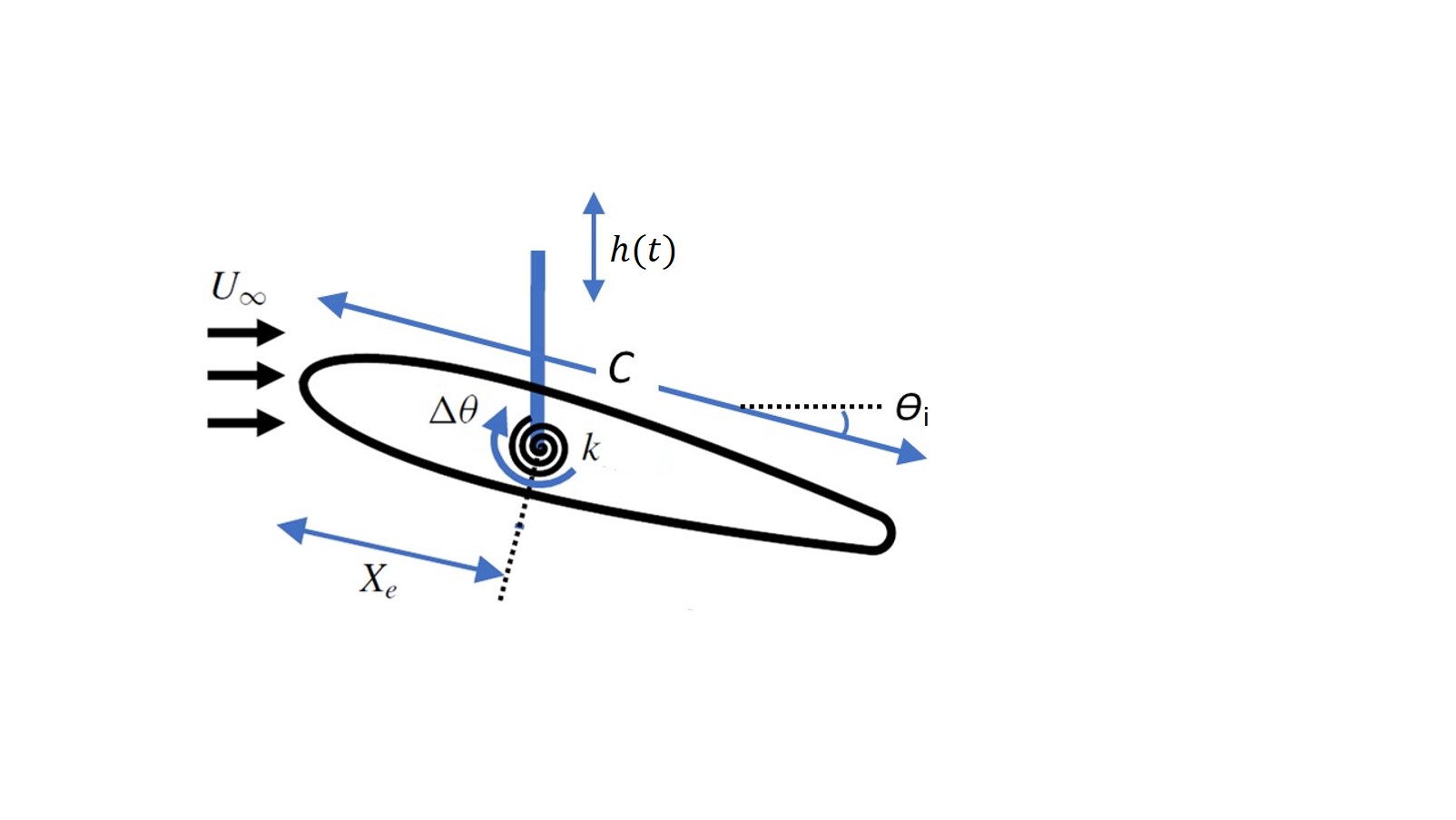}}
    \subfigure[]
    {\includegraphics[width=0.5\textwidth,trim={0cm 0cm 0cm 0cm},clip]{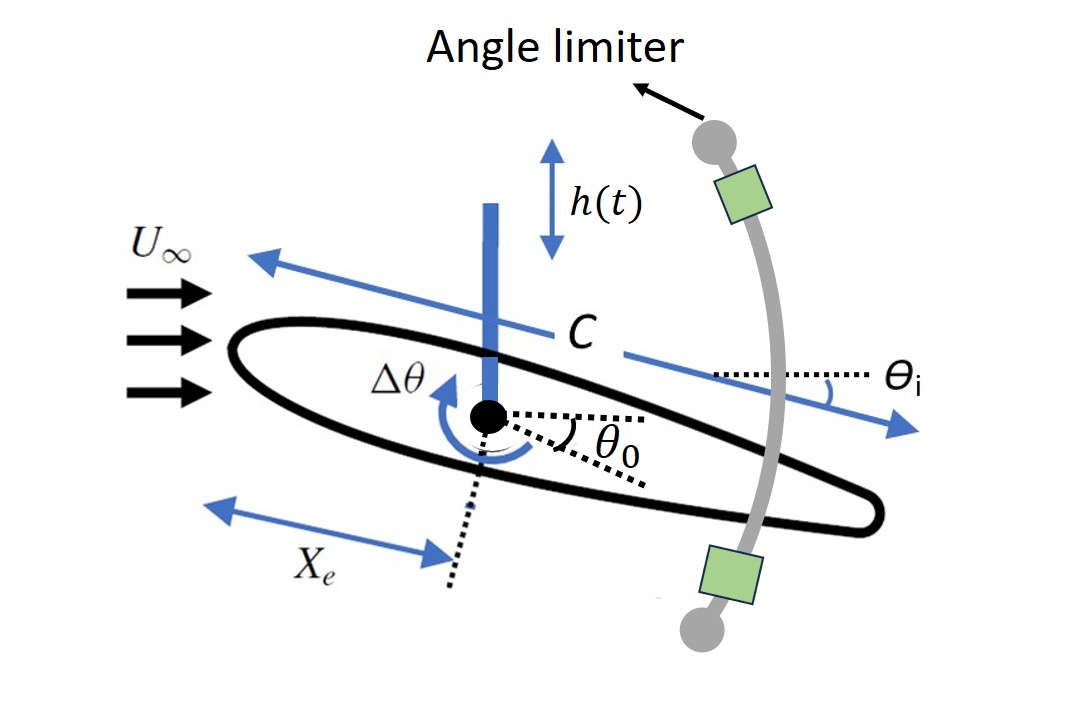}}
    \subfigure[]
    {\includegraphics[width=0.8\textwidth,trim={0cm 0cm 0cm 0cm},clip]{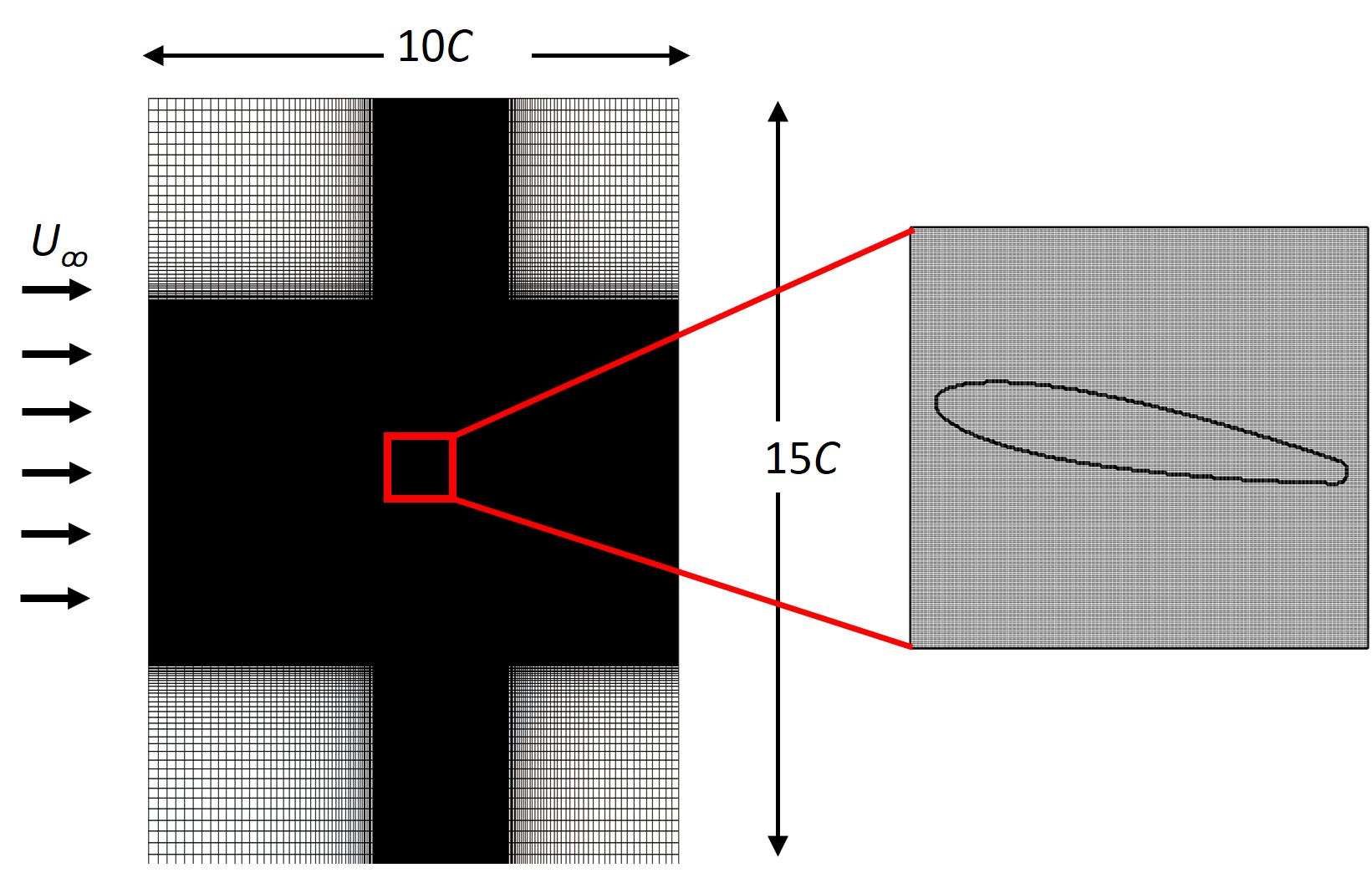}}\label{computational_domain}
    \caption{Schematic of the hydroelastic system used in this study for (a) spring-limiter and (b) angle-limiter. (c) Computational domain and close-up of the Cartesian computational grid.}
    \label{schematic}
\end{figure}

In the initial phase of the current study, we focus on a NACA0015 hydrofoil with a slightly rounded trailing-edge, which has been the subject of our previous work \citep{raut2024hydrodynamic,raut2025dynamics}. 
The governing incompressible Navier-Stokes equations expressed in the dimensionless form are as follows:
\begin{align}
    \frac{\partial \textbf{u}}{\partial t} + \textbf{u}.\nabla\textbf{u} &= -\mathbf{\nabla p} + \frac{1}{Re} \nabla^2 \textbf{u} \\
    \nabla . \textbf{u} &= 0
\end{align}
where $Re=\rho U_{\infty}C/\mu$ represents the chord-based Reynolds number. 
The hydrofoil undergoes the heaving motion with a prescribed amplitude which is connected with the given sea-state. In the case of a foil with a torsional spring, pitching is induced through the balance between flow-induced forces and restoring torque by the torsional spring, mounted at the hinge location, illustrated in Fig. \ref{schematic} (a). In the case of the angle-limiter, the foil is allowed to pitch freely due to the flow-induced forces but a constraint is imposed on the maximum allowable pitch angle. This may be implemented simply by an angle-limiter as illustrated in Fig. \ref{schematic} (b).

The linear torsional spring has a spring constant $k_\theta$, and the rotational hinge, or the pivot axis, is at a non-dimensional chordwise location of $X_e^{*} = X_e/C$. The maximum angle allowed by the angle-limiter is denoted as $\theta_{0}$. The equilibrium angular position of the spring is denoted by angle $\theta_i$ which has been kept zero throughout all the simulations in this study. The prescribed heaving in the $y$-direction is described by the equation $h(t)=h_0 \sin(2\pi f_h t)$, expressed in non-dimensional form as $H^*=H^*_0 \sin(2 \pi \textrm{St}_C t^*)$ with $\textrm{St}_C=f_h C/U_\infty$, $t^*=tU_\infty/C$ and $H^*=h/C$. In the case of spring-limiter, the flow-induced pitching motion is governed by the equation of a forced angular spring-mass oscillator. The equation is non-dimensionalized by the characteristic variables used for the flow (length: $C$; time: $C/U_{\infty}$), and the resulting non-dimensionalized equation governing the hydrofoil's pitch dynamics is given by:
\begin{equation}
    I^* \ddot{\theta}+k^*\theta  = C_M
    \label{solid_equation}
\end{equation}
where $C_M = M/(0.5 \rho U^2_{\infty} C^2 b)$ is the coefficient of pitching moment, and
$I^*=2I/(\rho C^4 b)$, $k^*= 2k/(\rho U^2_{\infty} C^2 b)$ are the dimensionless moment-of-inertia and spring stiffness respectively where, $b$ is the spanwise width of the foil. $C_M$ and $I^*$ are calculated with respect to $X_e^*$. The moment of inertia $I$ is calculated using the following volume integral $\rho\int_V r^2 dV$ where $r$ is the radial distance from the pivot location $X_e$. In this work, the torsional spring stiffness is expressed in terms of a non-dimensional frequency ratio $f_{\theta}/f_h$ where $f_{\theta}=(1/2\pi)\sqrt{k/I}$ is the natural frequency of the spring-mass system and $f_h$ is the frequency of the heaving motion in the $y$-direction. The above equation can be written in terms of $f_{\theta}$ as
\begin{equation}
    \ddot{\theta}+\Bigg(2\pi \frac{f_\theta}{f_h} \textrm{St}_C\Bigg)^2\theta  = \frac{C_M}{I^*}
    \label{solid_eq2}
\end{equation}
In the case of the angle-limiter, the second term in Eq. \ref{solid_eq2} is omitted and the pitch velocity is set to zero if the maximum allowable angle $\theta_{0}$ is reached and the pitching moment $C_M$  is tending to push the foil beyond the maximum angle. The foil is again allowed to pitch as soon as $C_M$ changes in direction.

The above setup involves six governing parameters Re, $I^*$, $X_e^*$, $H_0^*$, St$_C$ and $f_{\theta}/f_h$ (or $\theta_{0}$) for spring (or angle) limiter, highlighting the intrinsic complexity of this configuration. While comparing the performance of spring and angle-limiter, we fix the values of $Re$, $I^*$, $\theta_i$ and $X_e^*$ to 10,000, 0.05-0.12, 0 and 0.1 respectively and explore the effect of the spring stiffness ($f_{\theta}/f_h$), the maximum angle allowed in the angle-limiter ($\theta_{0}$), and sea-states ($H_0^*$ and St$_C$).  
The selected Reynolds number is sufficiently high to induce robust vortex shedding, thereby driving flutter, yet remains low enough to permit resolved simulations without excessive computational costs. The chosen non-dimensional moment of inertia about the elastic axis ranging from 0.05 to 0.12 corresponds to a solid-to-fluid density ratio of 1.8 and 4.3, respectively. 
While exploring the effect of varying $I^*$  could be of considerable engineering interest, we opt for a fixed value throughout this study to streamline the number of independent parameters. Nevertheless, it's important to acknowledge that adjusting $I^*$ is akin to altering the natural frequency, given by $f_{\theta} \sim 1/\sqrt{I^*}$ , which holds significance in this investigation. The pitch axis location $X_e^*$ is chosen to be 0.1 which was shown to generate the highest thrust in our earlier study \cite{raut2024hydrodynamic} on the single submerged flapping foil propulsion system. 

Three different sea-states are studied in this work corresponding to different wave heights and the underlying notion is that we would like the chosen propulsor design to provide thrust over a wide range of sea-states. The sea-states have been chosen based on the annual statistics of the North Atlantic ocean waves provided by Wang \emph{et al.} \cite{wang2019dynamic} and have been shown in Table \ref{tab_sea_states}. In determining the non-dimensional parameters, the chord length $C$ of the foil is set to 16 cm, a value that has been employed in the ``Wave Glider$^\text{®}$'' a system developed by Liquid Robotics (a Boeing company) which employs a wave-assisted flapping foil propulsion system. \cite{yang2018numerical}. Furthermore,  based on the data provided for the Wave Glider, the speed of the notional vehicle in the current study vehicle is taken to be 0.8 kt, 1.4 kt and 1.4 kt for sea-states 1, 2 and 3, respectively.  
\begin{table}[h!]
\centering
\begin{tabular}{>{\centering\arraybackslash}p{2.7cm}>{\centering\arraybackslash}p{3.2cm}>{\centering\arraybackslash}p{3.2cm}>{\centering\arraybackslash}p{1.5cm}>{\centering\arraybackslash}p{1.5cm}>{\centering\arraybackslash}p{1.5cm}}
\hline
\hline
 Sea-state \# & Wave height (m) &  Wave period (s) & $H_0^*$ & St$_C$ & St$_w$ \\
 \hline
 1 & 0.05 & 3.3 & 0.31 & 0.12 & 0.07\\ 
 2 & 0.25 & 5   & 1.56 & 0.04 & 0.12\\
 3 & 0.44 & 7.5 & 2.75 & 0.03 & 0.17\\ 
 \hline
\end{tabular}
\caption{Details of the sea-states used in the present study.}
\label{tab_sea_states}
\end{table}

\subsection{Computational Method}
The fluid-structure interaction in this study is simulated using the sharp-interface immersed boundary method solver ``ViCar3D,'' as described by Mittal \emph{et al.} \cite{mittal2008versatile} and Seo \& Mittal \cite{seo2011sharp}. This method is particularly well-suited for modeling flows around moving and deforming solid geometries, as it employs a simple body-non-conformal Cartesian grid that accurately captures the shape and motion of the immersed body. The approach maintains sharp interfaces around the immersed boundary, enabling accurate computation of surface quantities \cite{mittal2008versatile}. The flow equations are solved using a second-order fractional-step method, while the pressure Poisson equation is handled with the bi-conjugate gradient method. Spatial derivatives are discretized using second-order finite differences, and time-integration combines the second-order Adams-Bashforth and Crank-Nicolson methods. ViCar3D has been extensively validated and verified in prior studies involving stationary and moving boundary problems \cite{mittal2008versatile, seo2011sharp}.

The fluid-structure interaction model employs a loosely coupled approach, solving the flow and dynamic equations in a sequential manner. Hydrodynamic forces and moments are calculated at Lagrangian marker points on the surface of the solid body and are then applied to the solid dynamics equation (Eq. \ref{solid_equation}). The resulting angular velocity is imposed on the marker points. We employ a 2D hydrodynamic model for the WAP foils in our study. Our previous simulations \citep{raut2024hydrodynamic,raut2025dynamics} have established that intrinsic three-dimensionality does not significantly influence the thrust generation mechanism of these foils. We do not however consider finite-span effects, which likely diminish the thrust performance of all the foils. Finite-span 3D simulations are computationally expensive and it is not viable to do the O(100) simulations that we have done here to cover the large parameter space with finite-span models. The current simulations are therefore valid for large-span WAP foils such as those used in the Wave Glider, which has a foil aspect-ratio of 6.3\cite{kraus2012wave}. 

The hydroelastic system is simulated within a large computational domain measuring 10$C\times 15C$, with the leading hydrofoil positioned 4.5 chord-lengths downstream of the upstream boundary. The grid resolution around the solid body is isotropic, comprising approximately 250 points along the chord. Grid stretching occurs in all directions away from the rectangular region surrounding the foil and the near wake, resulting in a baseline grid of 641 $\times$ 1676 points. At the inlet boundary of the domain, a Dirichlet velocity boundary condition is applied, while zero-gradient Neumann conditions are specified at all other boundaries. Grid refinement and domain dependence studies, detailed in previous works \cite{raut2024hydrodynamic, raut2025dynamics}, affirm that the results obtained on this grid are well-converged and independent of the domain.

In all the simulations presented here, the foils are initially positioned at a zero pitch angle, located at $y=0$ (the mean heaving position), with zero initial angular velocity and heaving velocity. The equilibrium angle-of-attack for the hydrofoil is also set to zero. A constant free-stream flow is applied, allowing the dynamics to evolve naturally until the system attains a stationary state. For the cases simulated, achieving this stationary state typically requires approximately six oscillation cycles and the average quantities are calculated for several subsequent cycles.

\section{Results} 
%
\subsection{Foil Kinematics for Angle-Limiter and Spring-Limiter Mechanisms}
For the flow-induced motion simulations of the WAP system, a sinusoidal heaving motion is prescribed for the hydrofoil depending on the sea-state. The hydrofoil is allowed to pitch passively based on the pitching moment imposed on it by the flow and the spring/angle-limiter. Table \ref{tab_params} summarizes the parameters chosen for set of simulations that are carried out for the study of the pitch constraining mechanisms. Thus, a total of 24 simulations each for the spring-limiter and for angle-limiter have been performed to gain a comprehensive understanding of the effect of this mechanism on the thrust performance.

\begin{table}
\begin{tabular}{>{\centering\arraybackslash}b{2.7cm}  >{\centering\arraybackslash}b{5.9cm} >{\centering\arraybackslash}b{5.2cm} } 
 \hline
 \hline
 Sea-state \# & $\theta_0$ (deg.) (Angle-limiter) &  $f_\theta/f_h$ (Spring-limiter) \\
 \hline
 1 & 0, 1.25, 2.5, 3.75, 5, 10, 15, 20 & 1, 2, 3, 4, 5, 6, 7, 8 \\ 
 2 & 0, 2.5, 5, 7.5, 10, 12.5, 15, 20 &  3, 6, 9, 12, 15, 18, 21, 24   \\
 3 & 0, 2.5, 5, 7.5, 10, 12.5, 15, 20 & 3, 6, 9, 12, 15, 18, 21, 24 \\ 
 \hline
\end{tabular}
\caption{Parameters chosen for angle-limiter and spring-limiter to study pitch limiting mechanism.}
\label{tab_params}
\end{table}

\begin{figure}
    \centering
    \vspace{-0.5cm}
    \subfigure[  ]{\includegraphics[width=0.4\textwidth, trim={0cm 0cm 0cm 0cm}, clip]{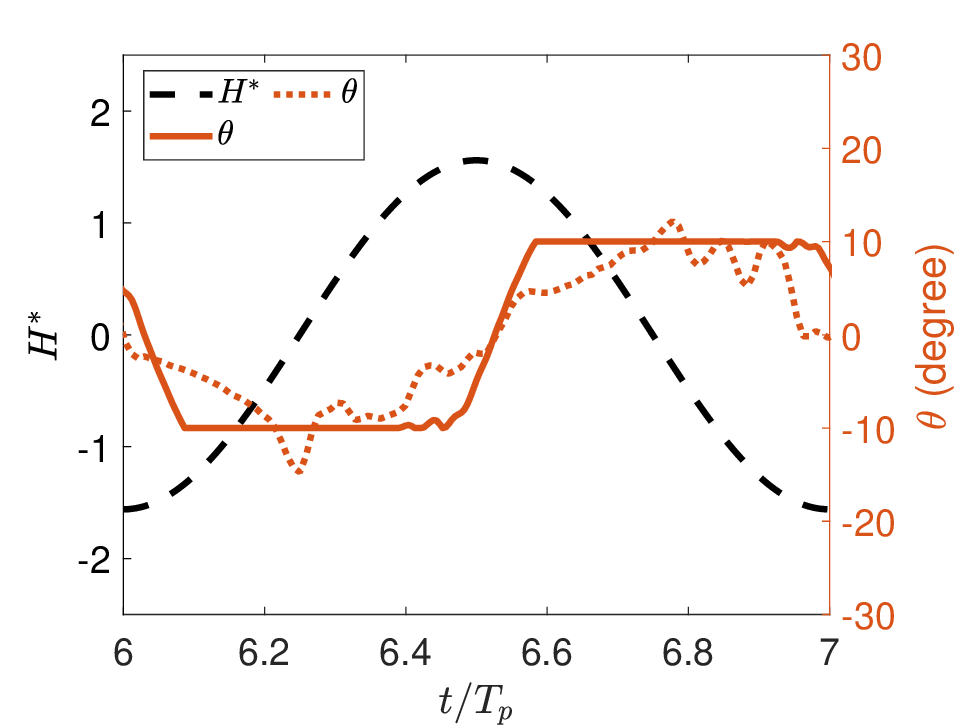}
    \label{compare_posi}}
    \subfigure[ ]{\includegraphics[width=0.42\textwidth, trim={0cm 0cm 0cm 0cm}, clip]{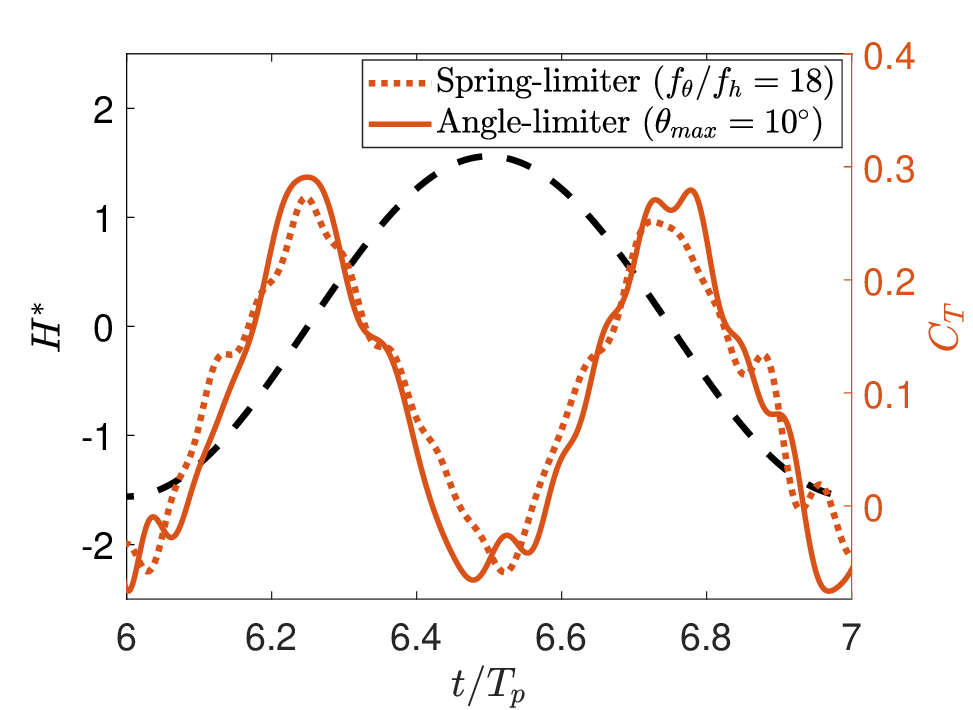}
    \label{compare_thrust}}\\
    \subfigure[ ]{\includegraphics[width=0.45\textwidth, trim={0cm 0cm 0cm 0.0cm}, clip]{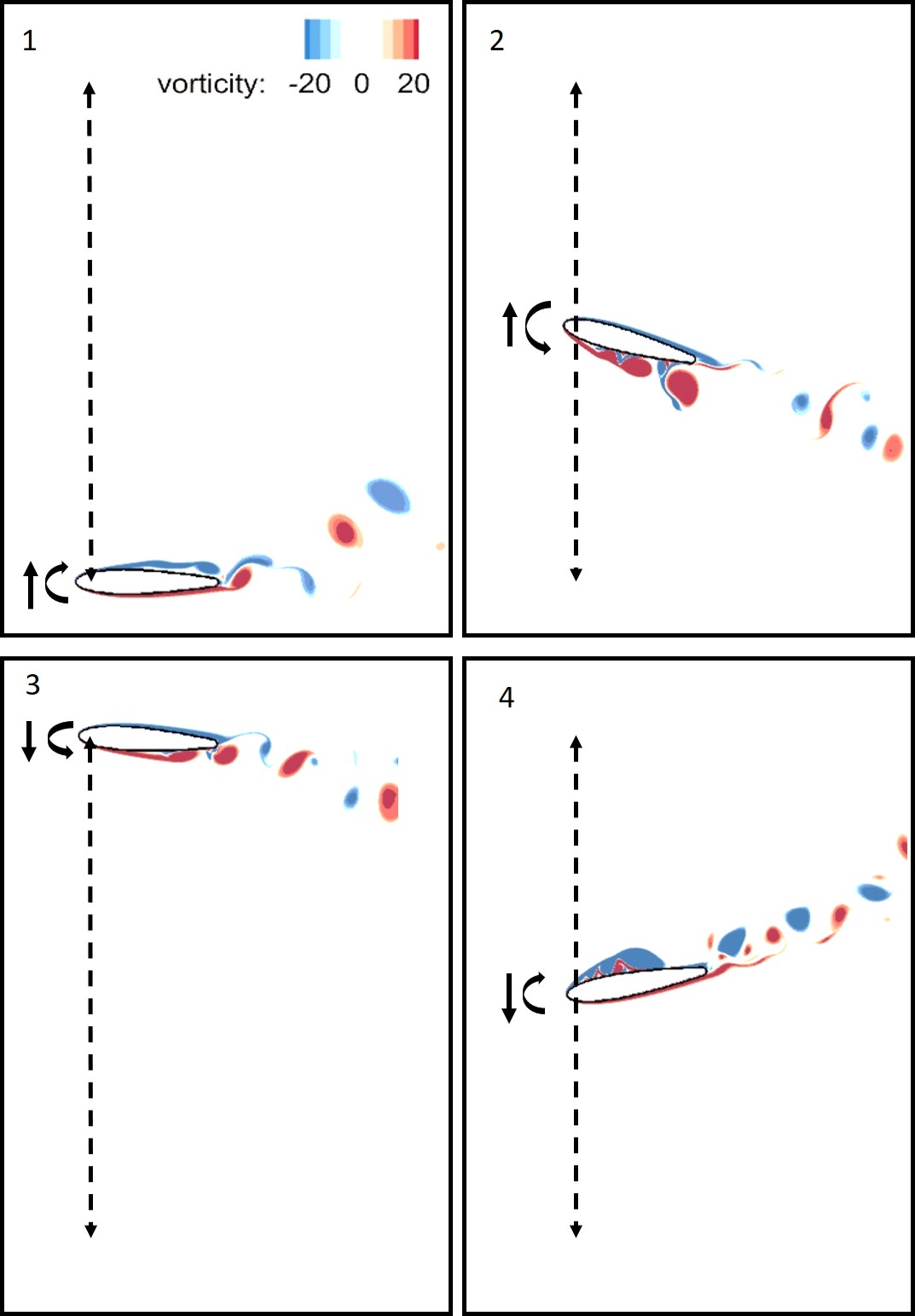}
    \label{torsional_spring_vort}}
    \hspace{1.0cm}
    \subfigure[ ]{\includegraphics[width=0.45\textwidth, trim={0cm 0cm 0cm 0.0cm}, clip]{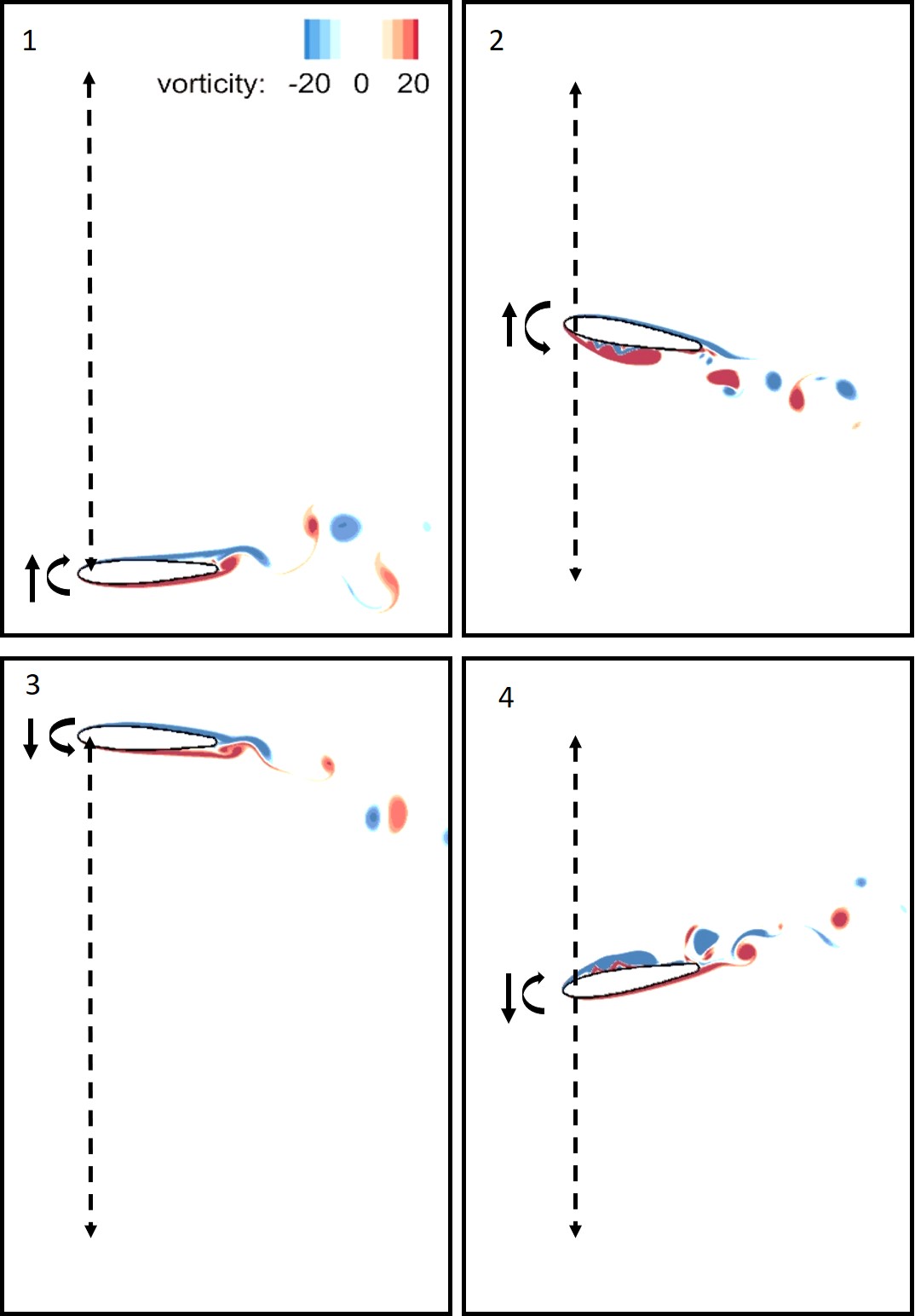}
    \label{angle_limiter_vort}}\\
    \caption{Comparison between the angle-limiter and the spring-limiter for the (a) pitching motion and (b) thrust coefficient for sea-state 2 with a pitching amplitude ($\theta_0$) of 10$^\circ$. (c) and (d) show the vorticity plots for spring-limiter and angle-limiter respectively.}
    \label{pass_pitch_rep}
    \vspace{1cm}
\end{figure}

We begin by presenting in Fig. \ref{pass_pitch_rep}, the results for the case with $\theta_{0}=10^\circ$ for the pitch-angle-limiter and $f_\theta/f_h=18$ for spring-limiter in sea-state 2. The frequency ratio ($f_\theta/f_h$) of 18 is chosen for this initial description because it generates a pitch oscillation amplitude of $9.84^{\circ}$ which is close to the maximum angle allowed by angle-limiter case ($\theta_{0}=10^\circ$). It should be noted that since the pitching motion of the foil with the spring-limiter exhibits complex behavior (\ref{compare_posi}), we calculate the pitch oscillation amplitude ($\theta_0$) for this case by taking a Fourier transform and measuring the amplitude of the dominant frequency ($f_h$). 
%

Figure \ref{compare_posi} shows the variation of heave and pitch for the foil with pitching motion controlled by the angle-limiter and the spring-limiter after the simulation has achieved a stationary state. We note that the angle-limiter has a simple pitching motion which resembles a sinusoid with a top-hat profile. On the other hand, the spring-limiter exhibits a complex pitch response that consist of rapid variations driven by the elastic recoil of the torsional spring. The time series for the coefficient of thrust ($C_T=-F_x/\frac{1}{2}\rho U^2_\infty C$) corresponding to the same oscillation cycle are shown in Fig. \ref{compare_thrust} for both the spring-limiter and the angle-limiter and we note that both cases have a nearly sinusoidal variation in this quantity with superposed higher frequency oscillations. 

Fig \ref{torsional_spring_vort} and \ref{angle_limiter_vort} shows the vortex shedding for these two cases. As the hydrofoil starts to heave in the upward direction, it experiences an increasing clockwise (negative) moment resulting in the pitching up of the foil. However, this pitch-up and heave-up motion increases the effective angle-of-attack of the foil and leads to the formation of leading-edge vortices (LEVs)  on the suction surface of the hydrofoil as seen in Fig. \ref{torsional_spring_vort}(2). These LEVs generate a suction pressure on the adjacent surface and this generates a positive thrust and a negative lift component. During the downstroke, the LEV formation process switches to the top surface resulting in a generation of a positive lift and a positive thrust (Fig. \ref{torsional_spring_vort}(4)). A similar process of LEV formation occurs in the case of the angle-limiter, as can be seen in Fig. \ref{angle_limiter_vort}. For these cases, with pitching amplitude of $10^\circ$, the thrust generated by both the mechanisms is similar and corresponds to an average thrust coefficient of 0.1.

\subsection{Angle-Limiter versus Spring-Limiter for Different Sea-States}
\begin{figure}
    \centering
    \vspace{-0.5cm}
    \subfigure[]{
    \includegraphics[width=0.47\textwidth, trim={0cm 0cm 0cm 0cm}, clip]{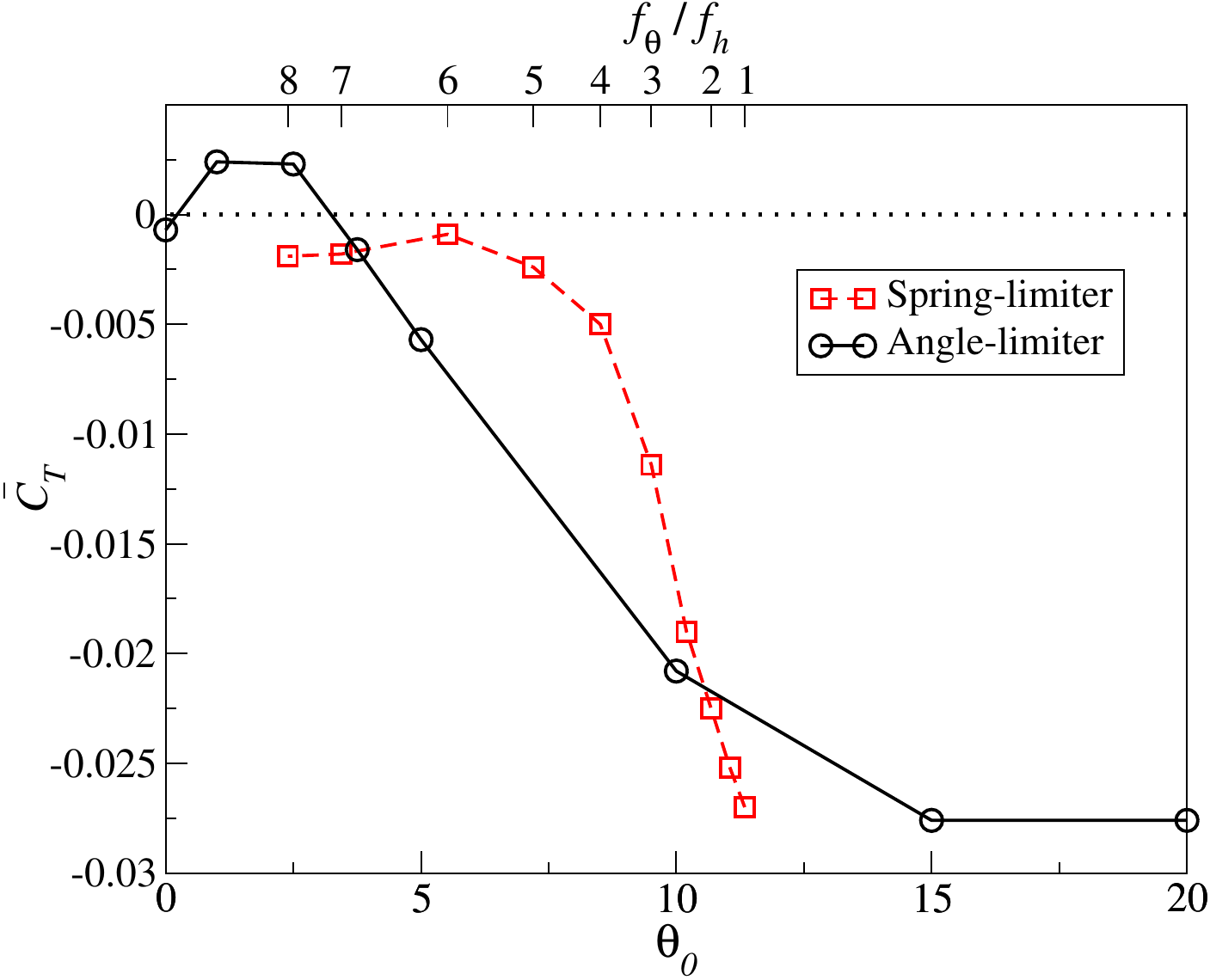}}
    \subfigure[]{
    \includegraphics[height=0.27\textheight,width=0.5\textwidth, trim={0.0cm 0cm 0cm 0cm}, clip]{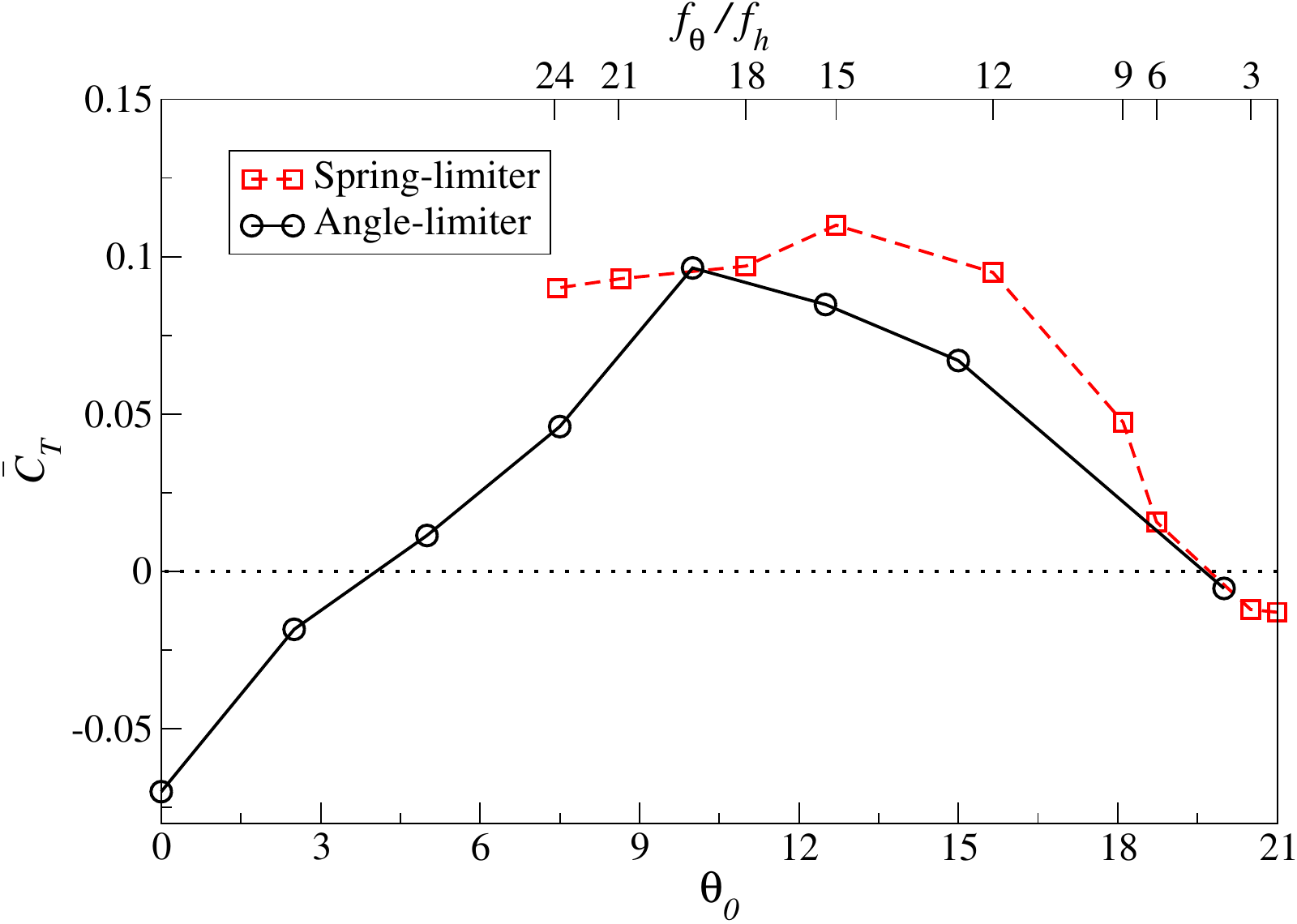}}    
    \subfigure[]{
    \includegraphics[height=0.27\textheight,width=0.48\textwidth, trim={0.0cm 0cm 0cm 0cm}, clip]{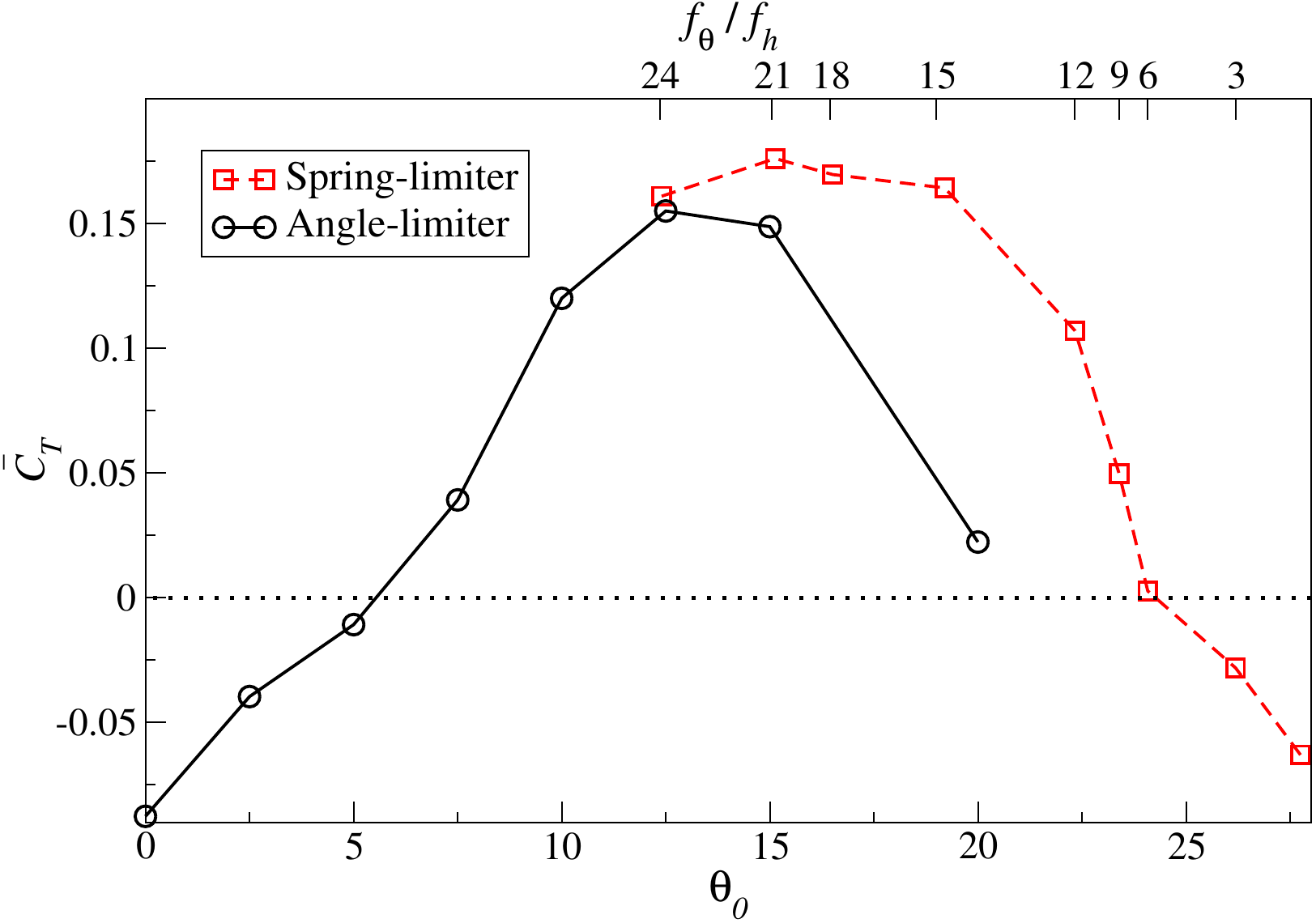}}    \caption{Average thrust coefficient ($\overline{C}_T$) over one oscillation cycle for angle-limiter and spring-limiter at different pitching amplitudes for (a) sea-state 1 ($H_0^*=0.31, \text{St}_C=0.12$), (b) sea-state 2 ($H_0^*=1.56, \text{St}_C=0.04$) and (c) sea-state 3 ($H_0^*=2.75, \text{St}_C=0.03$)}
    \label{ss1_thrust}
    \vspace{1cm}
\end{figure}
We next perform a comparative study between the spring-limiter and the angle-limiter designs in terms of thrust generation by varying the stiffness (or $f_\theta$) of the torsional spring and the maximum angle allowed by the angle-limiter ($\theta_{0}$) for different sea-states shown in table \ref{tab_sea_states}. Fig. \ref{ss1_thrust} (a) shows the average thrust generated by the angle-limiter ($\bar{C}_T$) in sea-state 1 for different pitch amplitudes ($\theta_0$). This is the most challenging sea-state for a WAP propulsor since the wave-height is small and indeed, we find that drag is generated for most of the values of $\theta_{0}$ that we have employed. A small amount of thrust is generated for lower values of $\theta_{0}$ with the highest thrust occurring at $\theta_{0}=2.5^{\circ}$. Fig. \ref{ss1_thrust} (a) also shows the same results for the spring-limiter with varying spring stiffness ($f_\theta/f_h$) which is shown on the upper $x$-axis. For this mechanism, drag is generated for \emph{all} the cases with lowest drag occurring at $f_\theta/f_h=6$ or $\theta_0=5.5^\circ$. 

\begin{figure}
    \centering
    \vspace{-0.5cm}
    \subfigure[]{
    \includegraphics[width=0.48\textwidth, trim={0cm 0cm 0cm 0cm}, clip]{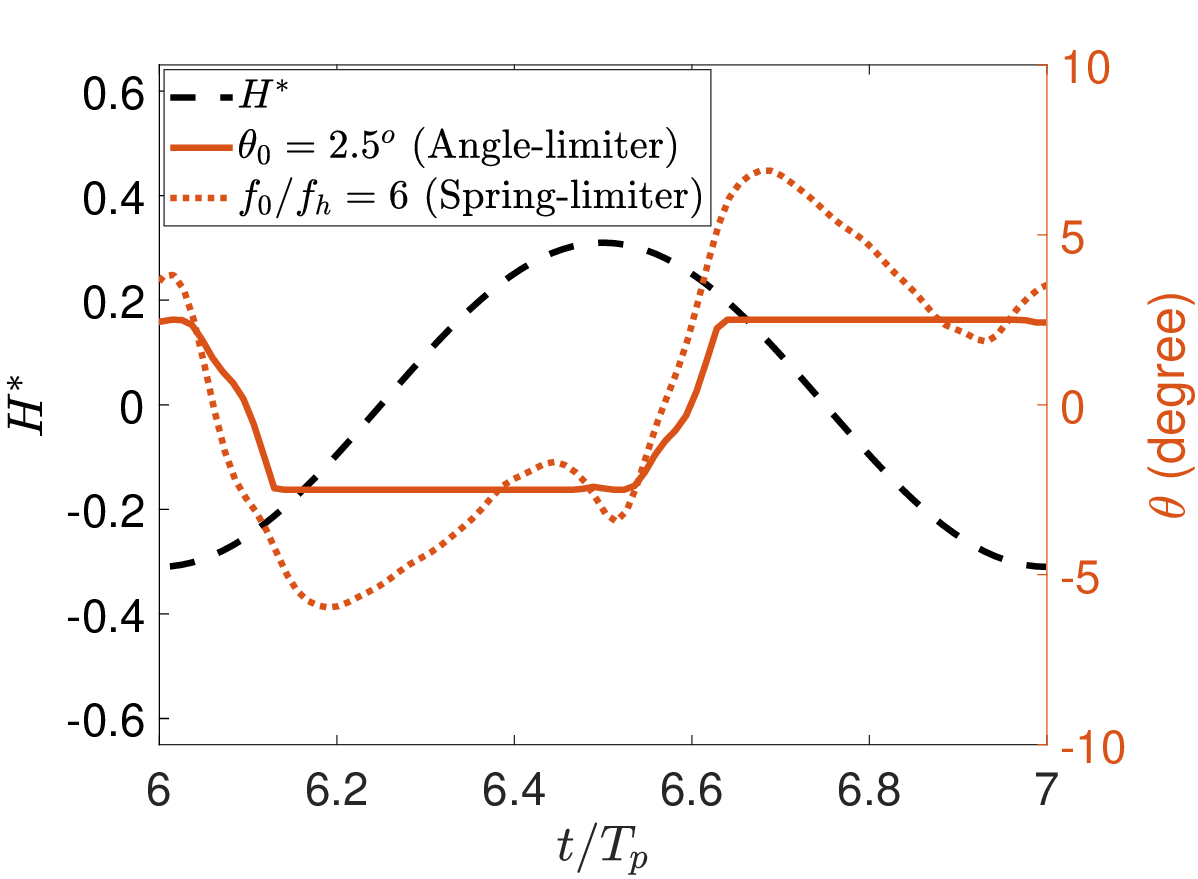}
    \label{ss1_thrust_posi}}
    \subfigure[]{
    \includegraphics[width=0.48\textwidth, trim={0.0cm 0cm 0cm 0cm}, clip]{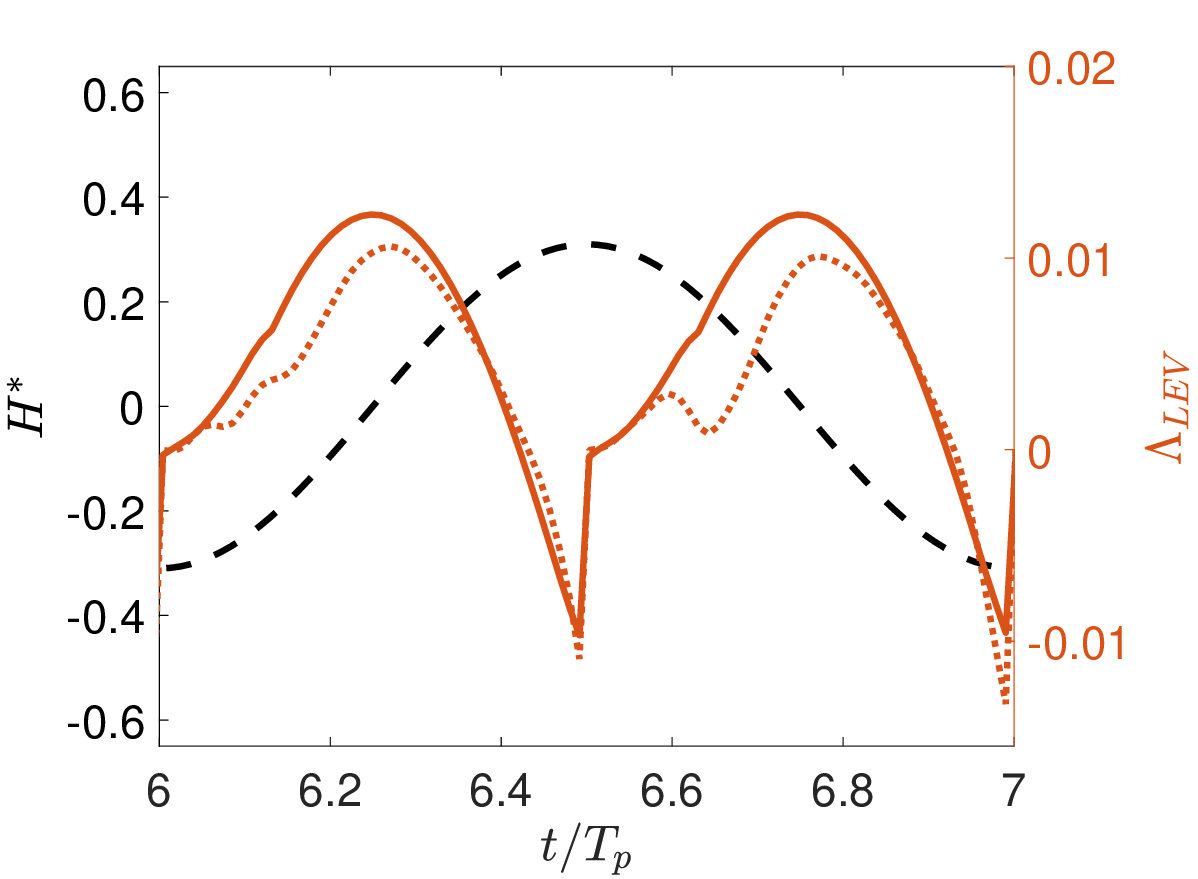}}    
    \label{compare_ss1}
    \caption{Comparison between spring-limiter and angle-limiter at peak thrust or low drag condition in sea-state 1.}
    \label{ss1_thrust_comp}
    \vspace{1cm}
\end{figure}

Fig. \ref{ss1_thrust_comp} (a) shows a comparison in the heaving and pitching motion for the angle-limiter at peak thrust ($\theta_o=2.5^\circ$) and the spring-limiter at the lowest drag condition ($f_\theta/f_h$) for sea-state 1. We note that while the angle-limiter by design, limits the pitch angle to a precise value, the spring-limiter has a more complex time-variation of pitch with the pitch angle reaching relatively high values. This occurs due to the non-linear interaction between the two dominant time-scales involved in the pitching motion in the case of the spring-limiter - the heaving frequency $f_h$, and the natural frequency of spring-mass system $f_\theta$. Fig \ref{ss1_thrust_comp} (b) shows the variation in the thrust factor  $\Lambda_\text{LEV}$ (see Eq. \ref{K_expanded} in the Appendix) over one periodic cycle for the heaving and pitching motion in Fig. \ref{ss1_thrust_comp} (a). We note that it is precisely at the phases where the pitch angle reaches large values when the $\Lambda_\text{LEV}$ for the spring-limiter is lower than that of the angle-limiter.  
Since $\Lambda_\text{LEV}$ is directly proportional to thrust coefficient (see \ref{eq_ctbar} in Appendix A), this clearly shows that the reason for the under performance of the spring-limiter mechanism is its highly unsteady pitch variation with large excursions in the pitch angle, which do not allow the LEV of high strength to be developed at phases where they can contribute most effectively to thrust.

Fig. \ref{ss1_thrust} (b) shows the average thrust coefficient over one oscillation cycle for sea-state 2 using both the angle-limiter and the spring-limiter. Compared to sea-state 1, the optimum value for pitch amplitude has shifted towards right with the optimum value for the angle-limiter occurring at $\theta_{0}=10^\circ$ and for spring-limiter at $\theta_0 = 12.4^\circ$. Similarly for sea-state 3 (Fig. \ref{ss1_thrust} (c)), the optimum value of pitch amplitude for the angle-limiter and the spring-limiter shifts toward right and now occur at $\theta_{0}=12.5^\circ$ and $\theta_0=15.05^\circ$ respectively. 
%
\begin{figure}
    \centering
    \vspace{-0.5cm}
    \includegraphics[width=0.52\textwidth, trim={0cm 0cm 0cm 0cm}, clip]{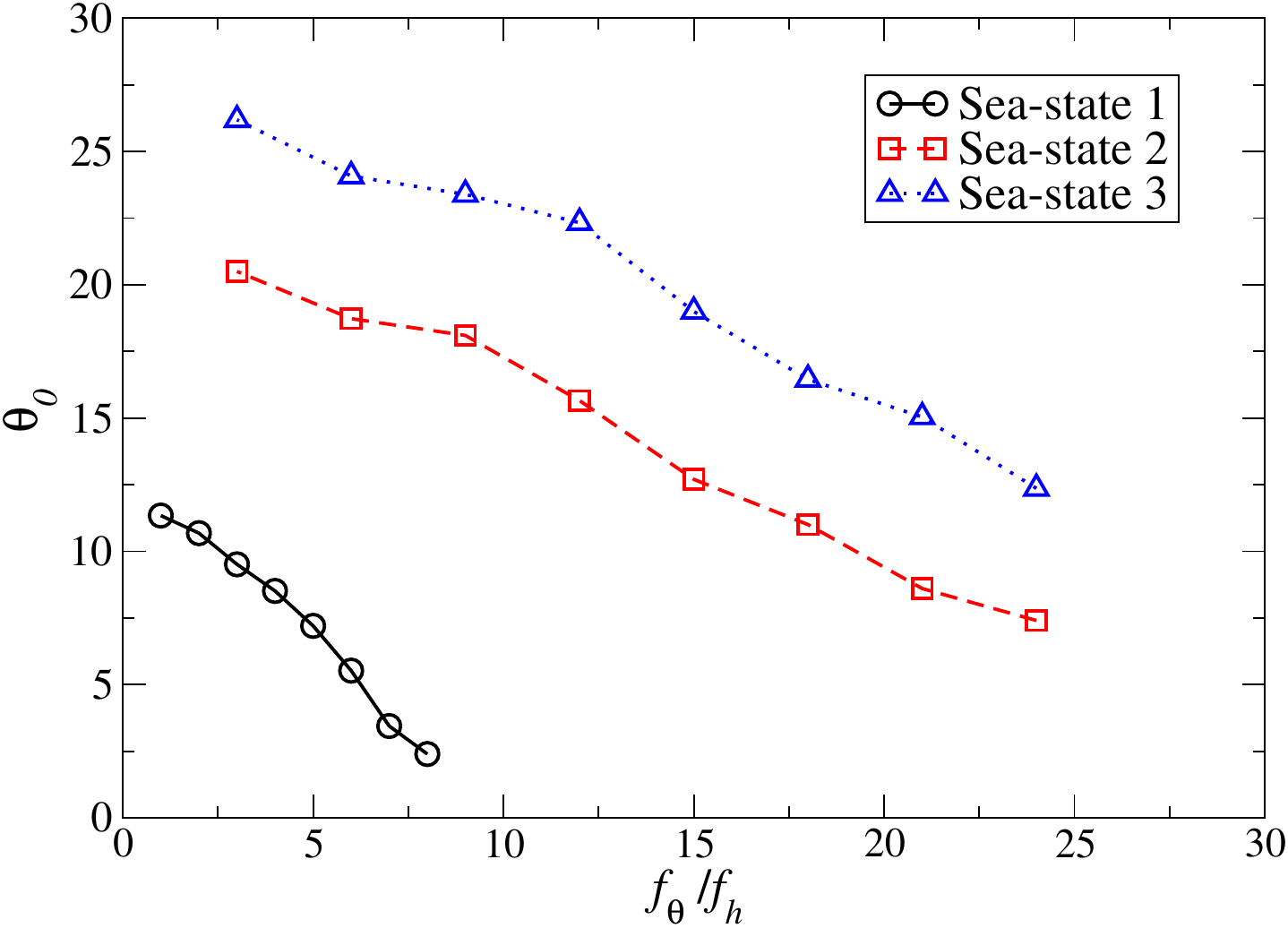}       
    \caption{Plot of pitch amplitude $\theta_0$ versus natural frequency of spring-mass system $f_\theta/f_h$ for all the sea states.}
    \label{freq_theta0}
    \vspace{1cm}
\end{figure}
For sea-state 1, the angle-limiter performs better as it generates a small amount of thrust for lower $\theta_{0}$ values and the spring-limiter generates drag for all values of $f_\theta/f_h$. For sea-state 2, the spring-limiter performs better generating 8.1\% more thrust than the angle-limiter. Similarly, for sea-state 3, the spring-limiter performs better generating 10\% more thrust than angle-limiter. Overall, it can be seen that the peak thrust generated by the mechanism of the spring-limiter and the angle-limiter is almost the same in all sea-states. However, in order to extract the maximum thrust the pitching amplitude needs to be adjusted for a given sea-state. In the case of the angle-limiter, this is done simply by adjusting the maximum angle allowed in the pitching motion. In the case of the spring-limiter however, this is done indirectly by adjusting the stiffness of the torsional spring (Fig. \ref{freq_theta0}). In Fig. \ref{freq_theta0} we plot the maximum pitch angle of the WAP foil against the normalized spring stiffness and we find that while there is a monotonic decrease in the maximum pitch angle with increasing spring stiffness, the behavior is not linear, and furthermore, this relationship also depends significantly on the sea-state. Thus, it is not trivial to determine a one-to-one relationship between the sea-state and the spring stiffness that could be used in a feed-forward manner to achieve ``optimal'' thrust conditions for various sea-states with the spring-limiter. In addition, even for optimal conditions, the spring-limiter gives non-smooth pitching motion (Fig. \ref{compare_posi} and \ref{ss1_thrust_posi}) which are non-conducive to thrust generation, especially at low sea-state conditions. 

\subsection{Foil Shapes that Enhance LEVs and Thrust}
\begin{figure}
    \centering
    \includegraphics[width=0.8\textwidth, trim={0.0cm 0cm 0cm 0cm}, clip]{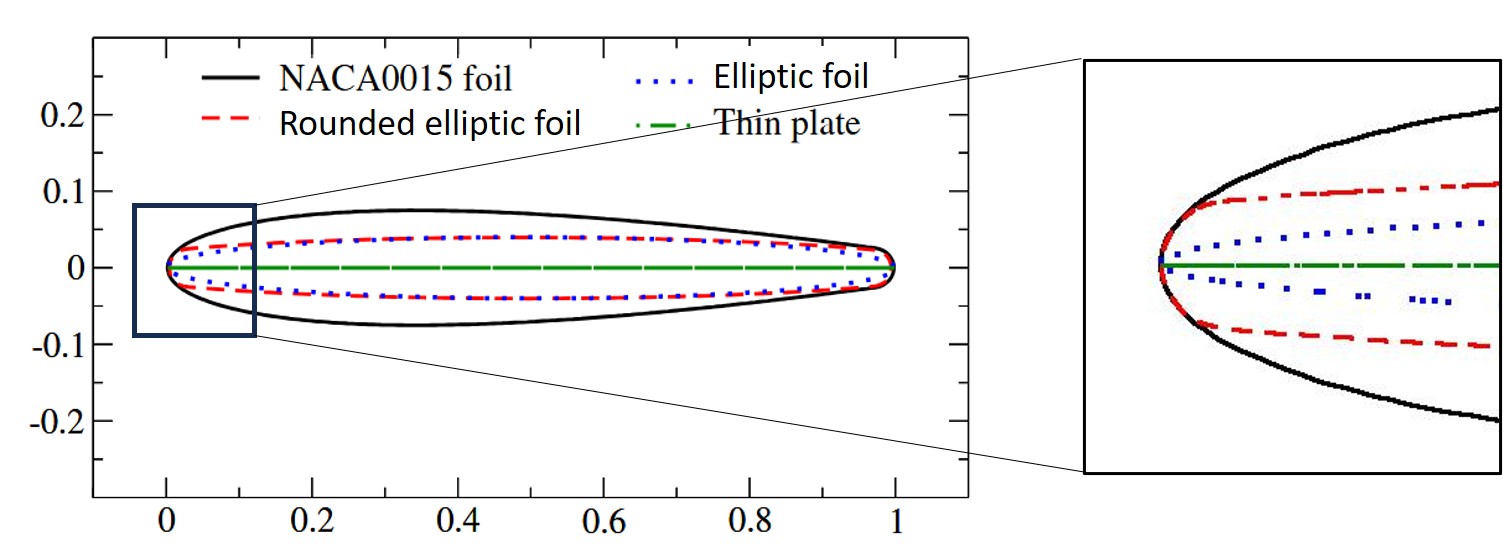}
    \caption{Plot of different proposed shapes of the foil in WAP system.}
    \label{comp_ell_re_mem}
\end{figure}
As has been seen in our previous work \cite{raut2024hydrodynamic} and in Fig. \ref{model_prediction}(b), the angle $\theta_s$, which is connected with the shape of the leading-edge, plays an important role in determining the performance of the foils given the dependence of thrust on the leading-edge vortex \cite{raut2024hydrodynamic, raut2025dynamics}. A thinner foil can also reduce form drag thereby leading to enhancement in the thrust. We therefore propose three shapes of the foil in addition to the NACA0015 foil as shown in Fig. \ref{comp_ell_re_mem} - these are  a 25:2 aspect-ratio ellipse, the same ellipse with rounded leading and trailing edges, and a thin flat plate. The two elliptic foils have a maximum thickness of 8\% and for the rounded ellipse, the radius of curvature of the leading and trailing edges is 2\% of the chord. It is worth noting that the rounded elliptic foil and the elliptic foil have the same cross-sectional area and they should therefore induce a similar form drag. The elliptic foils however has a sharper leading-edge and a comparison of these two cases should therefore highlight the effect of the leading-edge shape on the thrust performance. For consistency, the non-dimensional moment of inertia of all the foils about the pivot axis ($X_e^*=0.1$) is kept at 0.12. For each sea-state, a total of 32 simulations are performed with 8 different $\theta_\text{0}$ ($\theta_\text{0} = 0, 2.5, 5, 7.5, 10, 12.5, 15, 20$) and for each  foil. A total of three sea-states are considered leading to a total of 96 simulations for comparing the performance of different foil shapes.

\begin{figure}
    \centering
    \subfigure[]{
    \includegraphics[width=0.48\textwidth, trim={0.0cm 0cm 0cm 0cm}, clip]{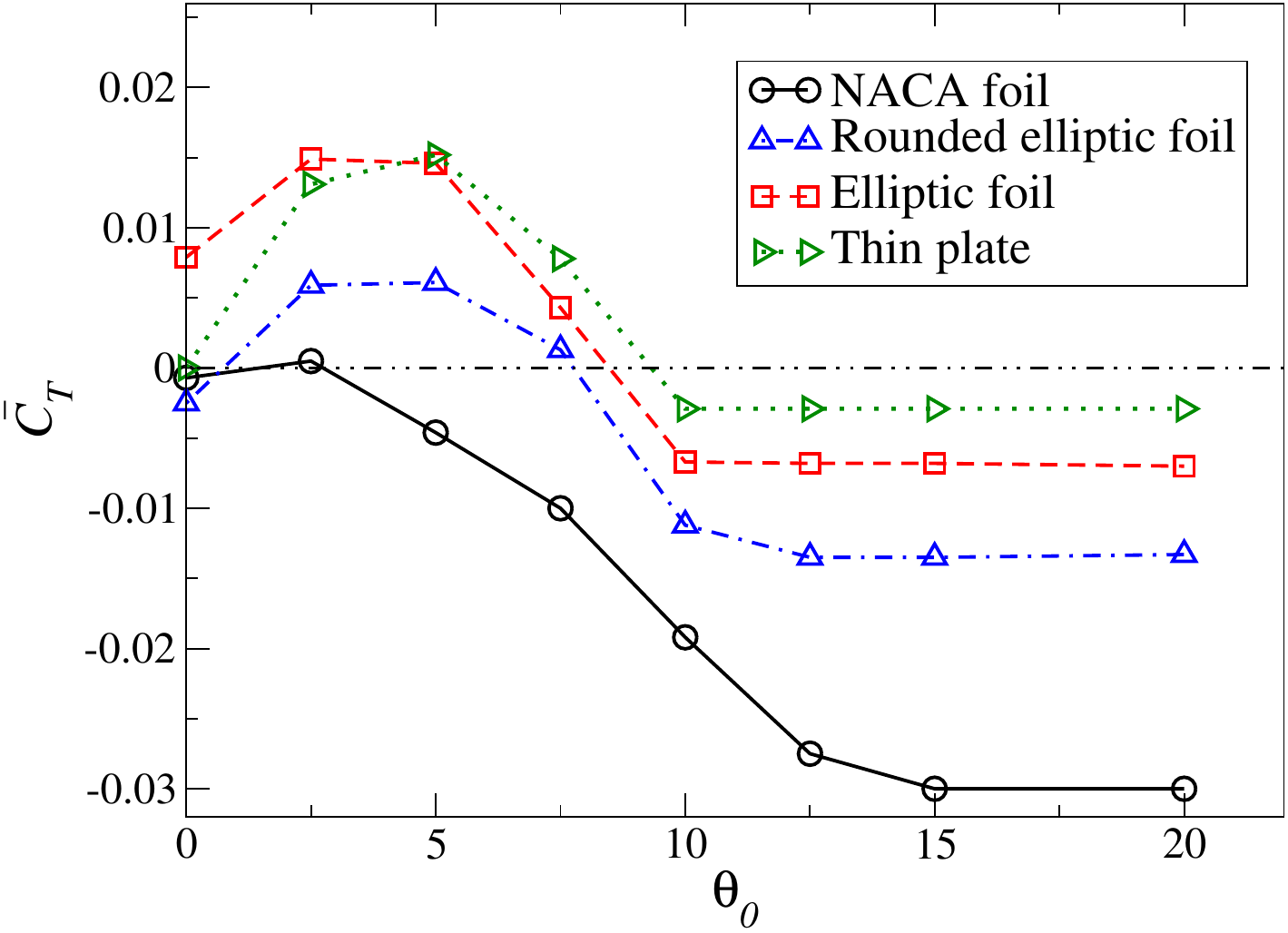}}
    \subfigure[]{
    \includegraphics[width=0.48\textwidth, trim={0.0cm 0cm 0cm 0cm}, clip]{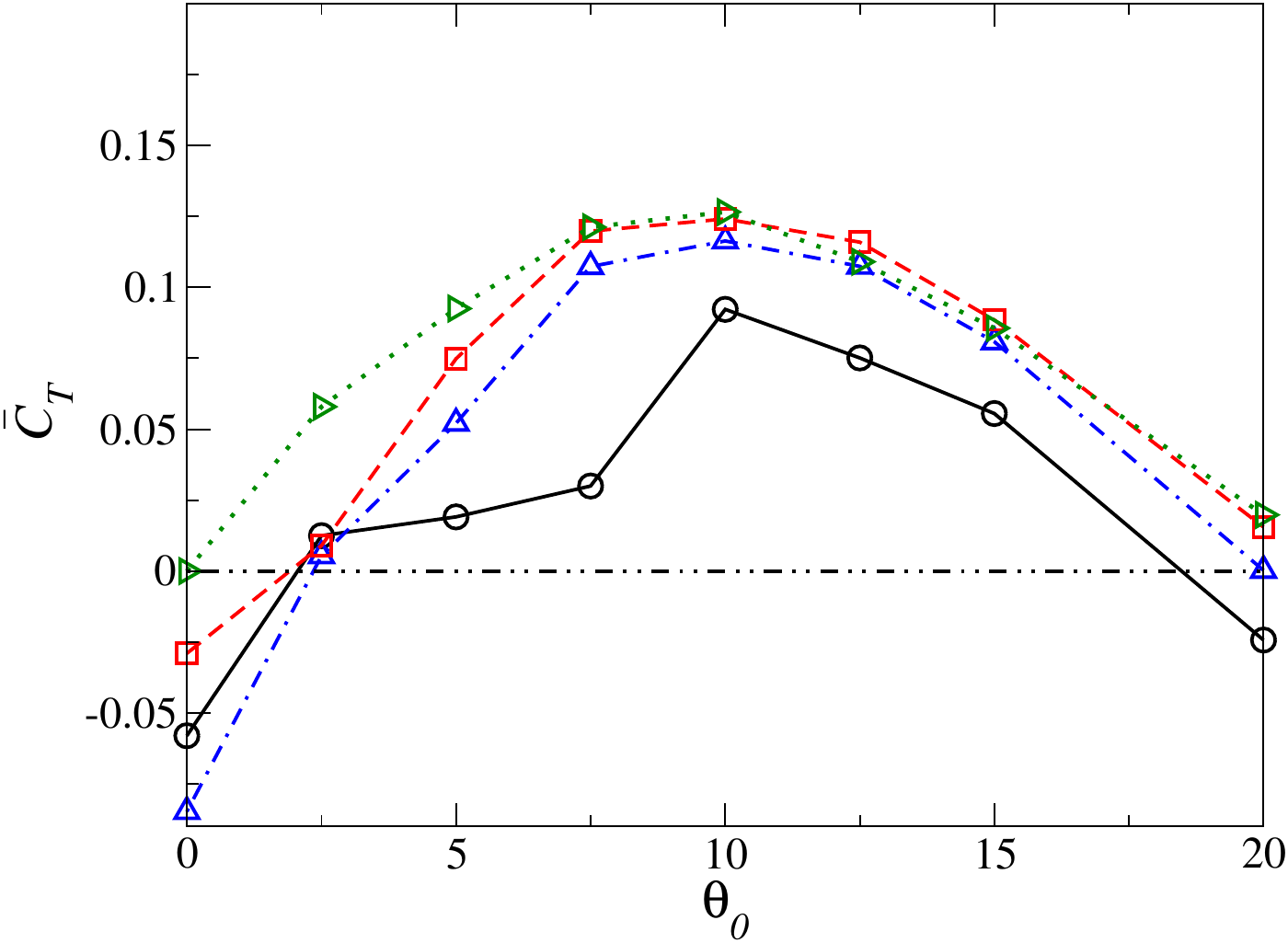}}
    \subfigure[]{
    \includegraphics[width=0.48\textwidth, trim={0.0cm 0cm 0cm 0cm}, clip]{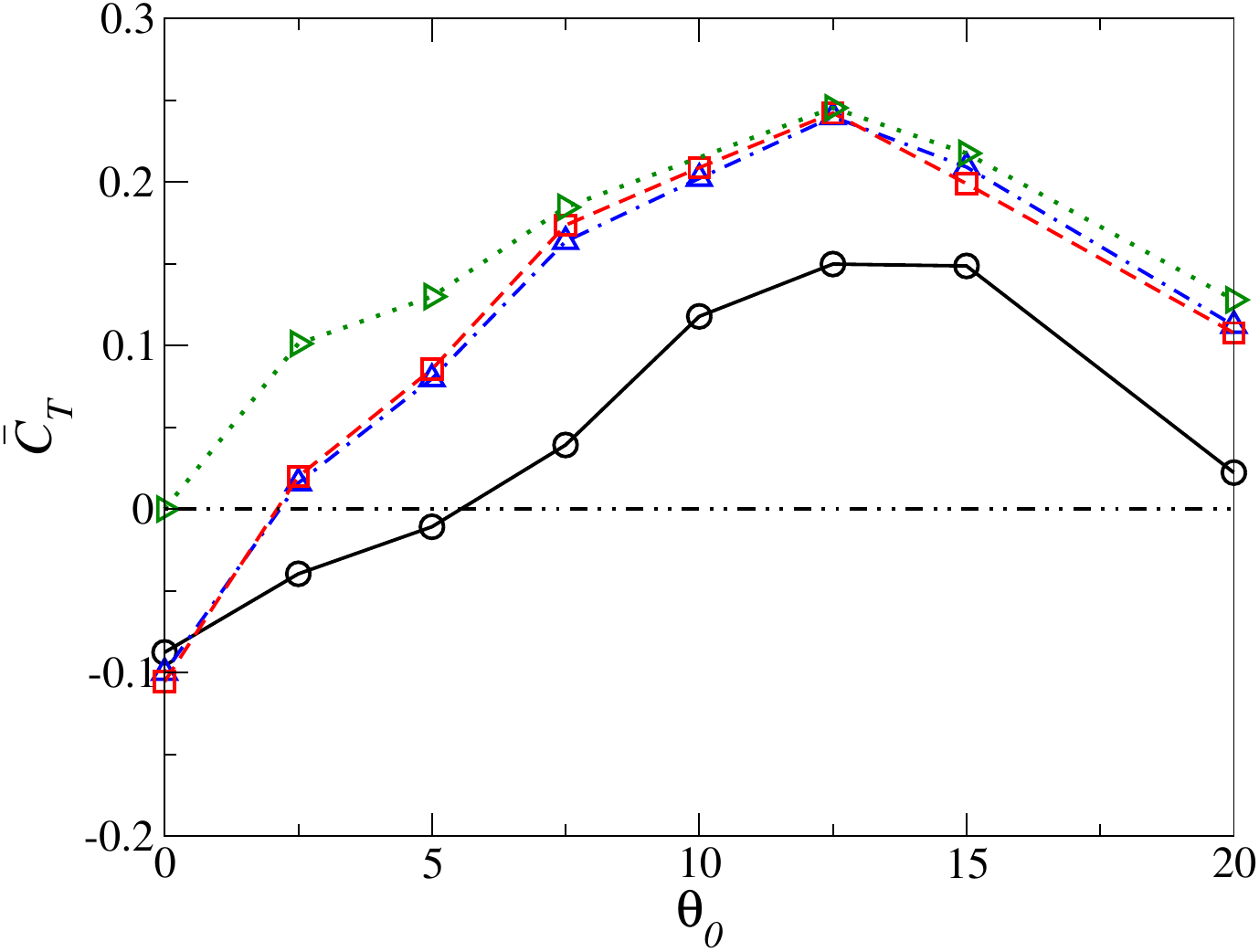}}
    \caption{Average thrust coefficient ($\bar{C}_T$) over one cycle for different shapes of the foil and different pitch amplitude ($\theta_0$) of angle-limiter for (a) sea-state 1, (b) sea-state 2 and (c) sea-state 3. }
    \label{thrust_comp_ell_re_mem}
\end{figure}
Fig. \ref{thrust_comp_ell_re_mem} shows the mean thrust coefficient for different shapes of the foil and for different sea-states. For sea-state 1, we notice significant differences in the performance of the foils and this is expected since for low wave amplitudes, we expect the shape of the foil to play an important role. First we note that for zero pitch angle, the ellipsoidal foil is the only one to generate thrust. The thin flat plate cannot generate thrust for this conditions since it's surface normal (the direction in which surface pressure acts) does not have a component in the thrust direction and the foil can therefore not generate any pressure thrust. The NACA0015 foil does not generate net thrust at this condition either because it generates a large form drag (similar to a static foil) for this low sea-state and zero pitch angle. The rounded ellipse also does not generate net thrust for this condition since the form drag balances out any thrust generated by this foil. The elliptic foil is able to generate significant LEV based thrust for this condition since it's surface normal does have a component in the thrust direction, and it's streamlined shape also reduces form drag. 

As the pitch amplitude is increased for this sea-state, the thrust for all the foils, except for the NACA0015 foil increases. The elliptic and  thin plate foils generate the largest thrust between a pitch amplitude of 2.5 and 5 degrees, but the thrust generated by the rounded elliptic foil is less than half that generated by the flat-plate and elliptic foils. This is clear proof that the sharp leading edge of these two foils facilitates the formation of a leading-edge vortex that enhances thrust at these low amplitude conditions. Except for the NACA0015 foil, all foils generate thrust for pitch amplitudes ranging from about $1^\text{o}$ to about $8^\text{o}$ for this sea-state. 

For sea-state 2, significant differences in mean thrust exist for the foils at low pitch-amplitudes. In particular, the thin plate foil generates higher thrust whereas the NACA0015 foil generates the lowest thrust. At pitch amplitude of $10^\text{o}$ the highest thrust is achieved, all foils except for the NACA0015 foil generate nearly equal thrust. Furthermore, the range of pitch amplitudes for which positive thrust is generated at this sea-state extends from about all foils generate thrust over pitch amplitude ranging from about $2^\text{o}$ to about $20^\text{o}$. The peak thrust generated at this sea-state is about an order of magnitude larger than that at sea-state 1.
The thrust performance for sea-state 3 is qualitatively similar to that for sea-state 2, except the peak thrust now occurs at a higher pitch-amplitude of about $12.5^\text{o}$ and peak thrust is nearly twice that for sea-state 2. In addition, positive thrust is generated over a much larger range of pitch amplitudes ranging from about  $2^\text{o}$ to beyond  $20^\text{o}$.

\begin{figure}
    \centering
    \subfigure[]{
    \includegraphics[width=0.48\textwidth, trim={0.0cm 0cm 0cm 0cm}, clip]{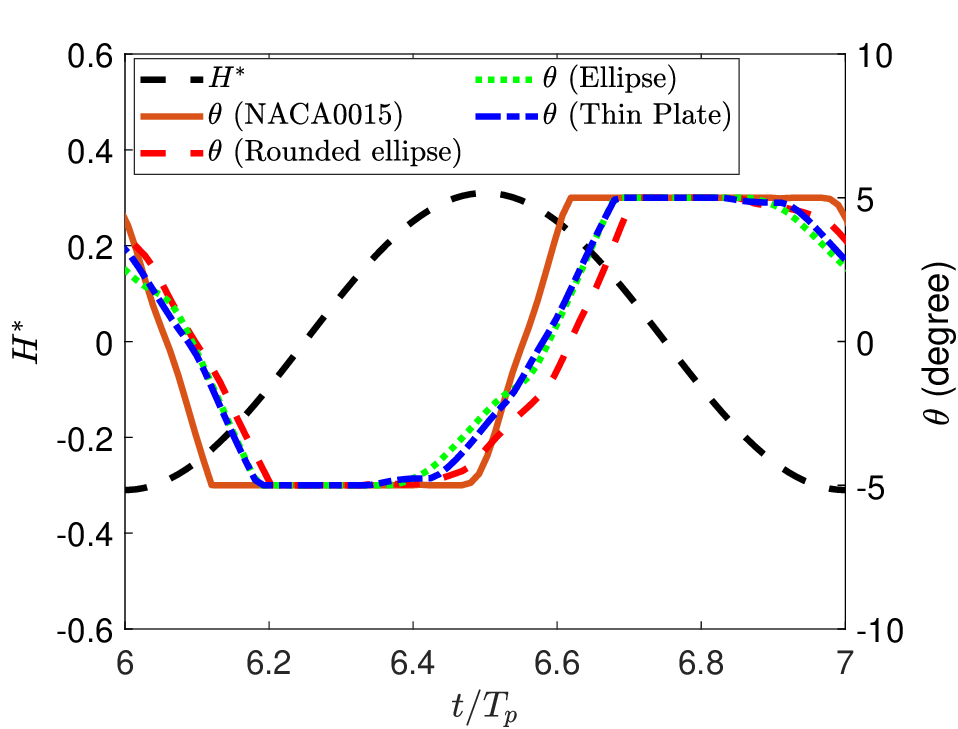}}
    \subfigure[]{
    \includegraphics[width=0.48\textwidth, trim={0.0cm 0cm 0cm 0cm}, clip]{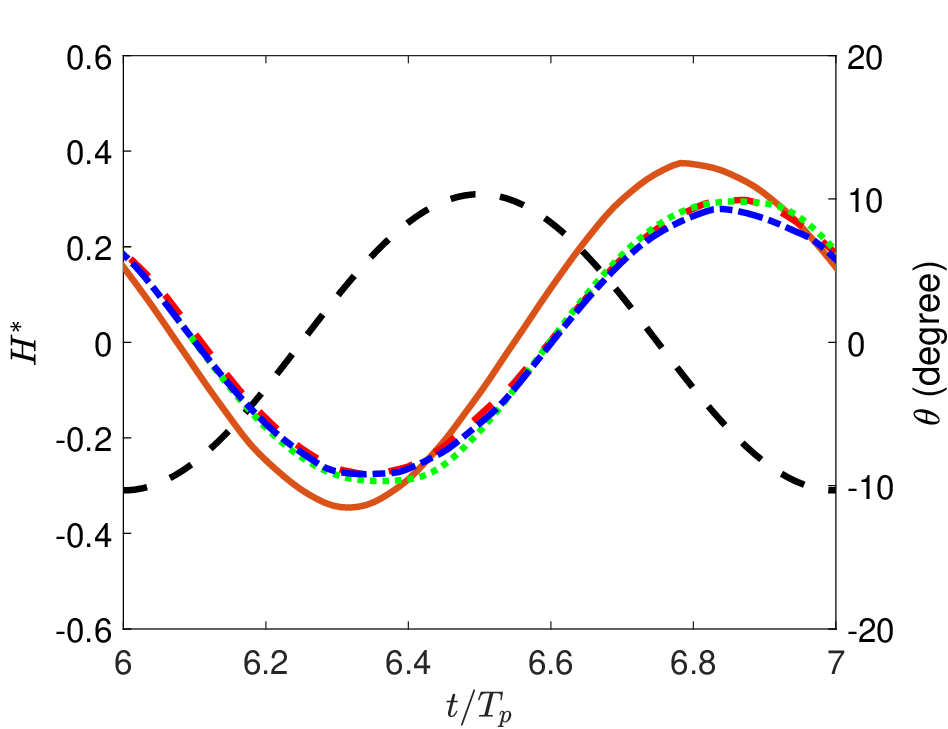}}
       \subfigure[]{    
    \includegraphics[width=0.48\textwidth, trim={0.0cm 0cm 0cm 0cm}, clip]{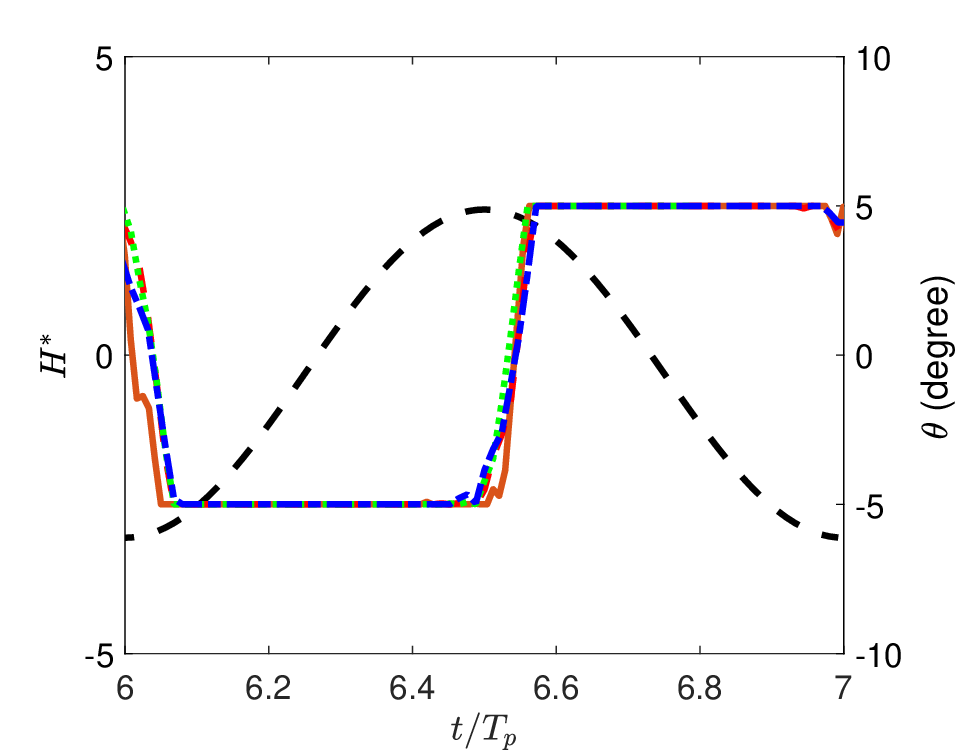}}
    \subfigure[]{    
    \includegraphics[width=0.48\textwidth, trim={0.0cm 0cm 0cm 0cm}, clip]{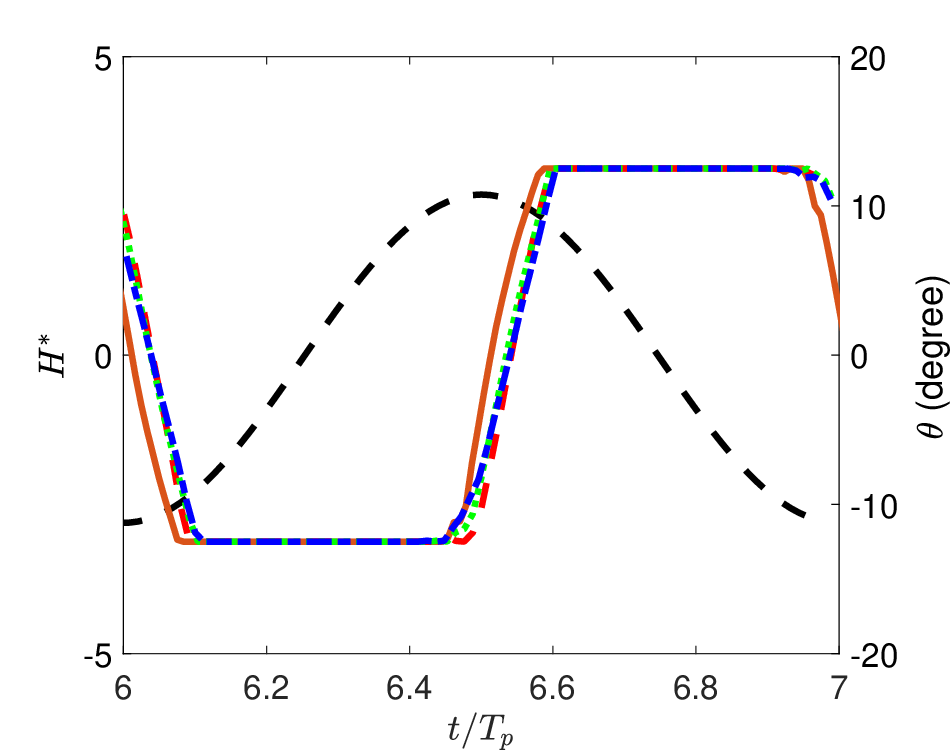}}
    \caption{Comparison of heaving and pitching motion for different shapes of the foil in (a,b) sea-state 1 and (c,d) sea-state 3 and with angle-limiter as the pitch constraining mechanism with (a,c) $\theta_0=5^\circ$ and (b,d) $\theta_0=12.5^\circ$.}
    \label{posi_comp_ell_re_mem}
\end{figure}
We now examine characteristics of the foil dynamics and the flow for these various foil shapes to gain a better understanding of the thrust performance. Fig. \ref{posi_comp_ell_re_mem} shows the time variation of the pitching motion of foils for two sea-states (1 and 3) and two different pitch amplitudes. we note that the NACA0015 foil stands out in terms of its motion among the four foils. The difference is particularly large for sea-state 1 (Fig. \ref{posi_comp_ell_re_mem} (a, b)) and diminishes for sea-state 3 (Fig. \ref{posi_comp_ell_re_mem} (c, d)). Since the moments-of-inertia are the same for all the foils, any difference in the motion is therefore primarily due to differences in the flows generated by these foils. 

\begin{figure}
    \centering
        \subfigure[]{
    \includegraphics[width=0.22\textwidth, trim={0.0cm 1cm 1cm 0cm}, clip]{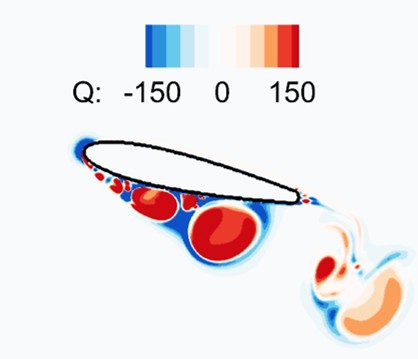}}
    \subfigure[]{
    \includegraphics[width=0.22\textwidth, trim={0.0cm 0cm 0cm 0.0cm}, clip]{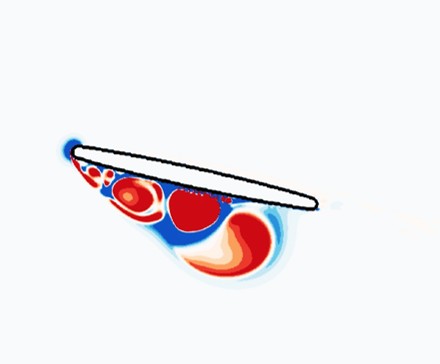}}
    \subfigure[]{
    \includegraphics[width=0.23\textwidth, trim={0.0cm 0cm 0.0cm 0.0cm}, clip]{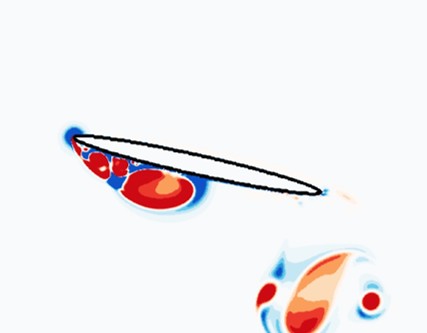}}
    \subfigure[]{
    \includegraphics[width=0.23\textwidth, trim={0.0cm 0cm 0cm 0cm}, clip]{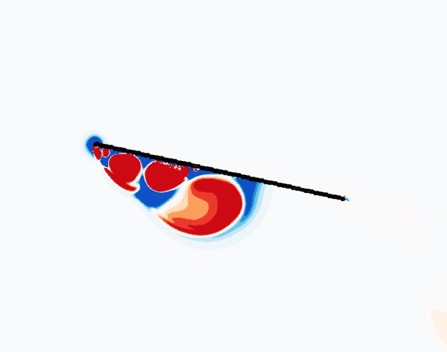}}\\    
    \subfigure[]{
    \includegraphics[width=0.22\textwidth, trim={0.0cm 1cm 1cm 0cm}, clip]{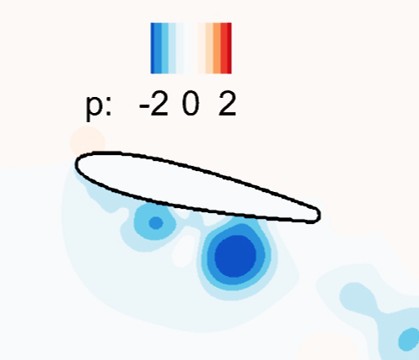}}
    \subfigure[]{
    \includegraphics[width=0.21\textwidth, trim={0.0cm 0cm 0cm 0.0cm}, clip]{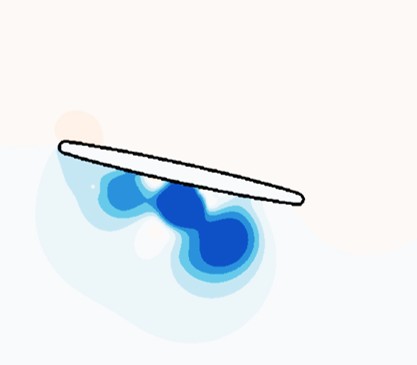}}
    \subfigure[]{
    \includegraphics[width=0.23\textwidth, trim={0.0cm 0cm 0.0cm 0.0cm}, clip]{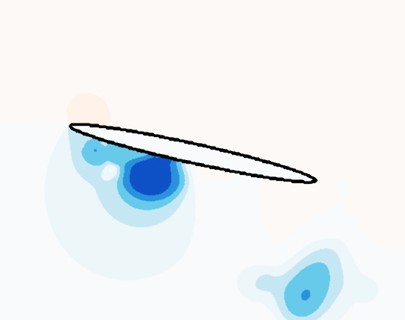}}
    \subfigure[]{
    \includegraphics[width=0.23\textwidth, trim={0.0cm 0cm 0cm 0cm}, clip]{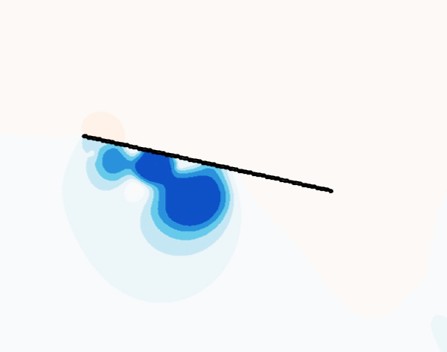}}
    \caption{Contours of Q (a-d) and pressure field (e-h) for sea-state 3. (a,e) NACA0015 foil, (b,f) Rounded ellipse, (c,g) Ellipse and (d,h) Thin plate at $t/T_p=6.25$ for sea-state 3 and $\theta_0=12.5^\circ$}
    \label{pres_comparison}
\end{figure}
Fig. \ref{pres_comparison}(a-d) shows a comparison of the $Q$-field for all the four foil shapes at mid-heave location for sea-state 3 and $\theta_0=12.5^\circ$. This corresponds to the peak thrust condition for all these foils. We note that there are significant differences in the vortex structures for the different foils, which are primarily due to the leading-edge shape, since the movement of the all the foils is very similar (see Fig. \ref{posi_comp_ell_re_mem} (d)). We note that the NACA0015 and the rounded elliptic foil, foils that do not have a sharp leading edge, have similar vortex structures and quite different from the other two foils, which have sharp leading edges. In particular, while the first two have vortex structures that stay in the vicinity of the foil even downstream of the center-chord, the vortex structures of the sharp leading-edge foils only stay attached till about the mid chord and separate beyond that. This has implications for thrust generation because the vortices are highly correlated with regions of low pressure (see Fig \ref{pres_comparison} (e-h)) and the suction pressure induced by the vortices is the primary causal mechanism for the thrust force. For the foils with curved surfaces (i.e. the NACA0015, rounded elliptic and elliptic foils), the first 50\% chord of the foil has a larger component of the surface normal in the thrust direction as compared to the second 50\% of the foil. Thus, the detachment of the vortices beyond 50\% chord for the elliptic and flat-plate foils allows the foil to generate relatively larger thrust force while diminishing the accompanying lateral forces. For the NACA0015 foil, this is is exacerbated by the fact that this foil has the weakest vortex near the leading-edge in terms of the induced suction pressure, and this results in this foil having the lowest thrust among all the 4 cases. 

\begin{figure}
    \centering
    \subfigure[]{
    \includegraphics[width=0.22\textwidth, trim={0.0cm 0.5cm 1cm 0cm}, clip]{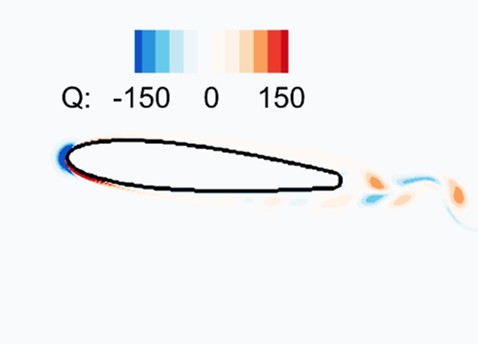}}
    \subfigure[]{
    \includegraphics[width=0.21\textwidth, trim={0.0cm 0cm 0.9cm 0.0cm}, clip]{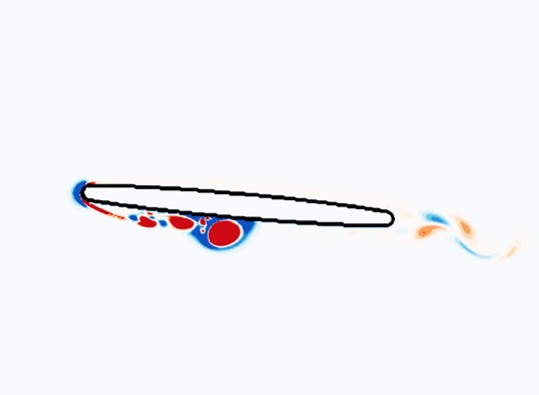}}
    \subfigure[]{
    \includegraphics[width=0.23\textwidth, trim={0.2cm 0cm 0.2cm 0.0cm}, clip]{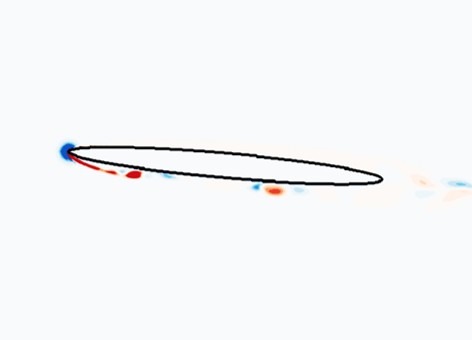}}
    \subfigure[]{
    \includegraphics[width=0.23\textwidth, trim={0.0cm 0cm 0.2cm 0cm}, clip]{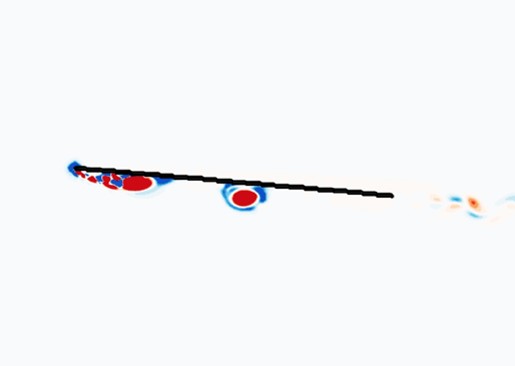}}\\
    
    \subfigure[]{
    \includegraphics[width=0.22\textwidth, trim={0.0cm 0cm 0cm 0cm}, clip]{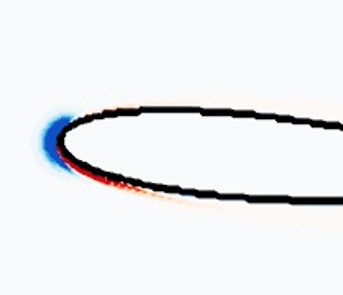}}
    \subfigure[]{
    \includegraphics[width=0.21\textwidth, trim={0.0cm 0.5cm 0.5cm 0.3cm}, clip]{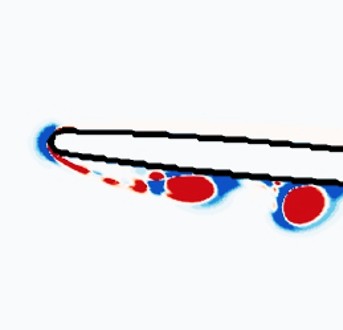}}
    \subfigure[]{
    \includegraphics[width=0.23\textwidth, trim={0.0cm 1.0cm 1cm 0.0cm}, clip]{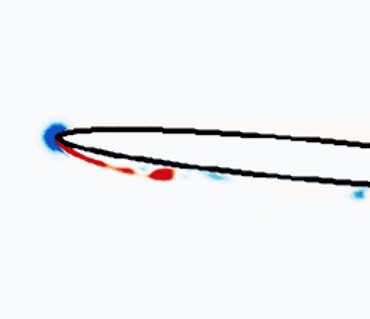}}
    \subfigure[]{
    \includegraphics[width=0.23\textwidth, trim={0.0cm 0.4cm 0.5cm 0cm}, clip]{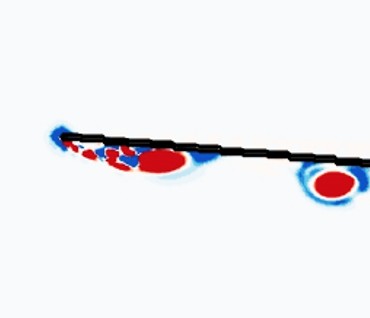}}\\

    \subfigure[]{
    \includegraphics[width=0.22\textwidth, trim={0.0cm 0.0cm 0cm 0cm}, clip]{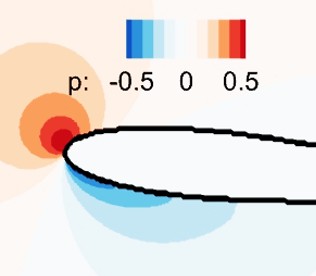}}
    \subfigure[]{
    \includegraphics[width=0.22\textwidth, trim={0.0cm 0.4cm 0.0cm 0.4cm}, clip]{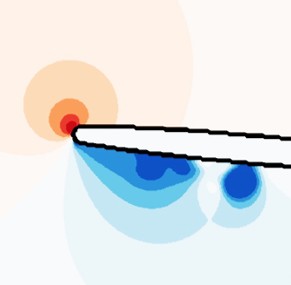}}
    \subfigure[]{
    \includegraphics[width=0.23\textwidth, trim={0.0cm 0.4cm 0cm 0.4cm}, clip]{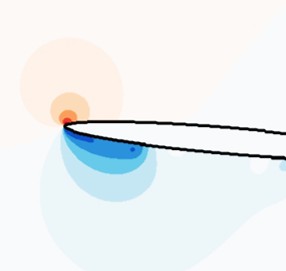}}
    \subfigure[]{
    \includegraphics[width=0.23\textwidth, trim={0.0cm 0.4cm 0.0cm 0.4cm}, clip]{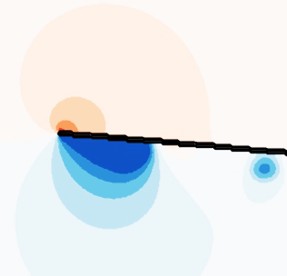}}\\
    \caption{Contours of Q (a-h) and pressure field (i-l) at $t/T_p=6.25$ for sea-state 1 and $\theta_0=5^\circ$. (a,e,i) NACA0015 foil, (b,f,j) Rounded ellipse, (c,g,k) Ellipse and (d,h,l) Thin plate. The second row of plots are zoomed-in views of the region near the leading-edges.}

    \label{q_comparison}
\end{figure}
We turn our attention to the flow features for sea-state 1 since this case is the most challenging for thrust generation. Fig. \ref{q_comparison} shows the vortex structures (including a zoom-in view of the first half of the foil) for $\theta_o=5^\circ$ case where all the cases, except for the NACA0015 foil, generate thrust. We note that while the NACA0015 does not generate rolled up LEVs at this small pitch amplitude, all three of the thin airfoils exhibit LEV formation and these generate large magnitude of suction pressure on the lower surface  (Fig. \ref{q_comparison}(i)-(l)) that contributes to thrust. Consequently, these three thin foils generate a positive thrust at this operating condition whereas NACA0015 generate negligible thrust. We note that the thrust of the rounded elliptic foil is about 50\% that of the other sharp LE foils and this highlights the effect that the rounding of the LE has on the vortex formation. In addition, the blunt leading edge of this foil also generates a larger positive pressure in the stagnation region (Fig. \ref{q_comparison}(j)) of the foil compared to the elliptic and flat-plate foil, which contributes to increased form drag and reduced net thrust.

Taken together, these results for different foil shapes indicate that thin foils such as the elliptic and flat plate foils are preferable for WAP propulsion applications. While the generation of thrust at high sea states is relatively straightforward over a wide range of pitch amplitudes, achieving effective propulsion at low sea states, such as sea-state 1, emerges as a critical consideration in the design of these propulsors. In this regard, our results indicate a slight performance advantage of the thin elliptic foil over the thin flat plate, and we therefore propose the thin elliptic foil as the preferred shape for WAP propulsor foils.

While an angle-limiter mechanism that is able to adjust the pitch amplitude to the optimal value in a feedforward manner based on an estimate of the Strouhal number  (see Fig. \ref{schematic} (b)) would enable the WAP system to maximize thrust for a given sea-state, this requires a capability for sensing the wave frequency and amplitude, as well as vessel speed and also an actuation system that can adjust the pitch amplitude of the angle-limiter. A simpler system with a \emph{fixed} pitch amplitude might be preferable if the complexity of an adjustable pitch is unacceptable. The current results also provide guidance for such a fixed pitch-amplitude system. In particular, our simulations indicate that for such a system, the angle-limiter should be set to a small value of about $5^\text{o}$ since this value of pitch amplitude generates thrust across all the sea-states, and especially sea-state 1, where the generation of propulsion from wave motion is challenging. 

\section{\label{conc}Conclusions}
High fidelity flow simulations with coupled fluid-structure interaction modeling are used to explore the performance of wave-assisted propulsion systems. The objective of this work is to leverage our understanding of the flow physics of these flapping foil propulsors to propose and assess design changes that can increase thrust across a range of sea-states.  

In our earlier work, we demonstrated that the leading-edge vortex (LEV) is the dominant feature in thrust generation and showed that thrust from a LEV for a given sea-state can be maximized by limiting the pitch-amplitude of the flapping foil propulsor. This has motivated our exploration of new designs of WAP propulsors that can provide improved thrust performance compared to our previous study. In particular, we have compared the performance of two pitch control mechanisms- a spring-limiter and an angle-limiter, under three different sea-states. We find that while at high-sea states both mechanisms are able to generate similar thrusts, the angle-limiter is more effective in generating thrust at sea-state 1, which is a critical operating point for such systems. Combined with the simplicity of the angle-limiter mechanism relative to the spring-limiter, we conclude that the angle-limiter mechanism is the preferred mechanisms for these propulsion systems.

The importance of the leading-edge vortex for thrust generation also motivated us to examine the effect of leading-edge shape on the thrust performance of these foils. In particular, in addition to the NACA0015 foil, we also examined a 25:2 elliptic foil with and without a rounded leading and trailing edge, and a thin flat plate. Our results suggest that thin foils, particularly elliptical and flat plate types, are well suited for WAP applications. While effective thrust generation is achievable across various pitch amplitudes in higher sea states, propulsion in low sea states—especially sea-state 1—is more challenging and thus critical in propulsor design. The thin elliptical foil demonstrates a marginal performance advantage over the flat plate and is recommended as the preferred geometry for WAP foils.

A feedforward angle-limiter mechanism that adjusts pitch amplitude based on estimated Strouhal number could optimize thrust for a range of sea-state conditions. However, this approach necessitates real-time sensing of wave characteristics and vessel speed, along with a responsive actuation system. Alternatively, a fixed pitch amplitude offers a less complex solution, which might be preferable in some designs. Simulations indicate that a fixed pitch amplitude of about 5$^\circ$ provides thrust generation across all sea states, including sea-state 1, offering a practical design compromise for simpler WAP systems.

In conclusion, the study demonstrates how insights into the flow physics of flapping foils can lead to designs that can significantly improve the performance of these wave-assisted propulsion systems across a range of operating conditions.

\subsection*{Acknowledgements}
This work is supported by ONR Grants N00014-22-1-2655 and N00014-22-1-2770 monitored by Dr. Robert Brizzolara. This work used the computational resources at the Advanced Research Computing at Hopkins (ARCH) core facility (rockfish.jhu.edu), which is supported by the AFOSR DURIP Grant FA9550-21-1-0303, and the Extreme Science
and Engineering Discovery Environment (XSEDE), which is supported by National Science Foundation Grant No. ACI-1548562, through allocation number TG-CTS100002.

\subsection*{Declaration of interests}
The authors report no conflict of interest.
\subsection{Appendix - LEV Based Model}
\begin{figure}
    \centering
    \includegraphics[width=0.6\textwidth, trim={0.0cm 0cm 0cm 0cm}, clip]{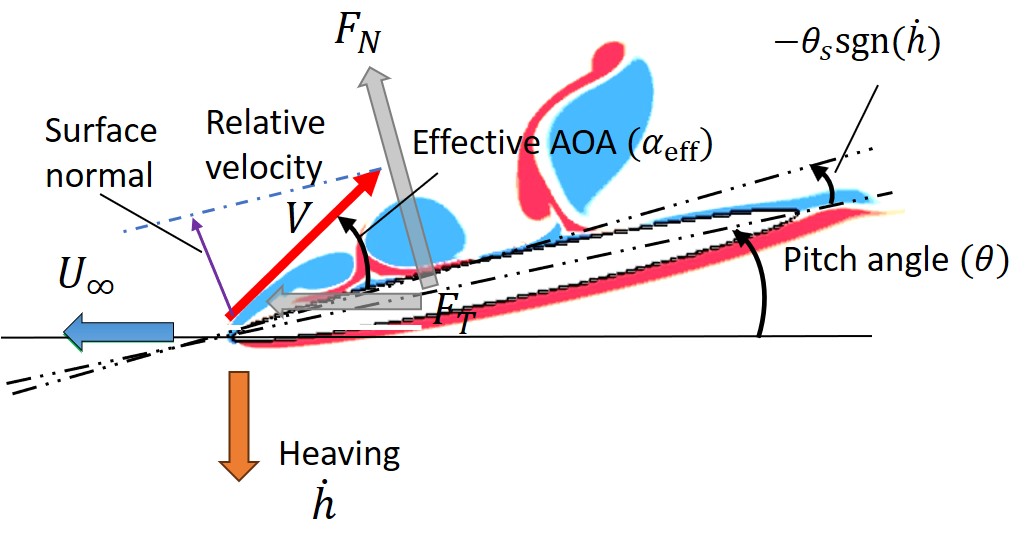}
    \caption{Schematic depicting the key features of the LEV based model.}
    \label{model_schema}
\end{figure}

In our previous work\cite{raut2024hydrodynamic}, we have we have shown the dominant role of LEVs in thrust generation for the WAP foils and have derived a new phenomenological model for LEV based thrust generation of flapping foils. We present the key aspects of the model here since it informs the design changes explored in this paper. As per the Kutta-Joukowski  theorem, the force perpendicular to the foil $F_N$(see Fig. \ref{model_schema}) is given by the following equation:
\begin{equation}
    F_N = \rho V \Gamma
    \label{F_N}
\end{equation}
here $\Gamma$ is the circulation of the hydrofoil and $V$ is the net relative velocity of the foil with respect to the flow given by $V=\sqrt{U^2_\infty+\dot{h}^2}$.  Our previous analysis has shown that the circulation $\Gamma$ is dominated by the leading-edge-vortex (LEV).  The formation and strength of the LEV is proportional to the component of velocity $V$ of the flow at the leading edge that is normal to the surface. This component is associated with the effective angle-of-attack ($\alpha _\textrm{eff}$) induced by the flow and the movement of the foil. Thus, $\Gamma$ can be expressed as 
\begin{equation}
 \Gamma \propto  V \sin (\alpha_\textrm{eff})   
 \label{circulation}
\end{equation}
where the effective angle-of-attack from the surface of the foil, which is given by
\begin{equation}
  {\alpha _\textrm{eff}} (t) = {\tan ^{ - 1}}\left({V_\textrm{LE}(t)/U_\infty} \right)- \theta_\textrm{LE}(t)
  \label{alpha_eff}
\end{equation}
where the first term represents the angle-of-attack induced at the leading-edge by the heaving and pitching of the foil with the lateral velocity of the leading-edge obtained as $ V_\textrm{LE}(t) = -\left(   \dot h- \dot{\theta}(t) X_e\cos{\theta(t)}\right)$. The second term $\theta_\textrm{LE}$ represents the inclination of the surface of the airfoil over which the LEV develops and induces it force on and is expressed as $\theta_\textrm{LE}(t) =\theta (t)- \theta_s \, \textrm{sgn}\left(\dot{h}(t) \right)$. In this expression, the variable $\theta_s$, is a fixed angle-offset between this surface and the chord of the airfoil and is related to the thickness and shape of the foil near the leading-edge.

Further assuming the following modified normalization of the normal force
\begin{equation}
    C_N = \frac{F_N}{\frac{1}{2}\rho V_\textrm{max}^2 C}
    \label{modified_thrust_coeff}
\end{equation}
where we employ the maximum value of $V$ instead of $U_\infty$ and using equation \ref{F_N}, \ref{circulation} and \ref{modified_thrust_coeff} we arrive at $C_N \propto \sin 
\left( \alpha_\textrm{eff} \right)$. From this it follows that the thrust coefficient satisfies the following proportionality
\begin{equation}
    C_T \propto \sin 
{\left( \alpha_\textrm{eff} \right)} \sin{\left( \theta_\textrm{LE} \right)}
    \label{eq_ctprop}
\end{equation}
Denoting the time-dependent function on the right-hand-side, which is written solely in terms of the kinematics and geometry of the foil by $\Lambda_\textrm{LEV}$, the mean thrust from this model can be expressed as
and the mean thrust coefficient can be expressed as 
\begin{equation}
    \bar{C}_T = K \bar{\Lambda}_\text{LEV} (\theta_s)
    \label{eq_ctbar}
\end{equation}
where $\Lambda_\text{LEV}(t;\theta_s)$ can be expanded (using Eq. \ref{alpha_eff}) as:
\begin{align}
    \Lambda_\text{LEV}(t;\theta_s) &=\sin 
{\left( \alpha_\textrm{eff} \right)} \sin{\left( \theta_\textrm{LE} \right)}=\sin {\left( {\tan ^{ - 1}}\left({V_\textrm{LE}(t)/U_\infty} \right)- \theta_\textrm{LE}(t) \right)} \sin{\left( \theta_\textrm{LE} \right)}\label{K_expanded}
\end{align}

For a given flapping foil, the above model for mean thrust has two unknowns - $K$, the constant of proportionality and $\theta_s$, which is connected with the shape of foil near the leading-edge. The values of these unknowns are determined for a NACA0015 foil undergoing combined pitching and heaving for a range of Strouhal numbers by using a least-square error fit for the 462 prescribed motion simulations that have been conducted for various pitch angle amplitudes, pitch axis locations and Strouhal numbers. The fit (see Fig. 16 in Raut \emph{et al.}\cite{raut2024hydrodynamic}) results in $K=3.41$ and $\theta_s = 3.62^\circ$. The fit has a $R^2$ value of 0.91 indicating a relatively high degree of correlation between the data and the model.

Furthermore, as shown in our previous work  \cite{raut2024hydrodynamic} the above model  predicts that thrust is maximized for a pitching amplitude that is given by 
\begin{equation}
    \theta^\text{opt}_{0} = 0.5 \tan^{-1}\left( \pi \text{St}_w \right)-\theta_s
    \label{theta0max_eq}
\end{equation}
Thus, control of the pitching amplitude is important for the foil to generate maximum thrust and the shape of the leading edge, which affects $\theta_s$, will also impact the thrust performance.


\begin{figure}
    \centering
    \includegraphics[width=0.7\textwidth]{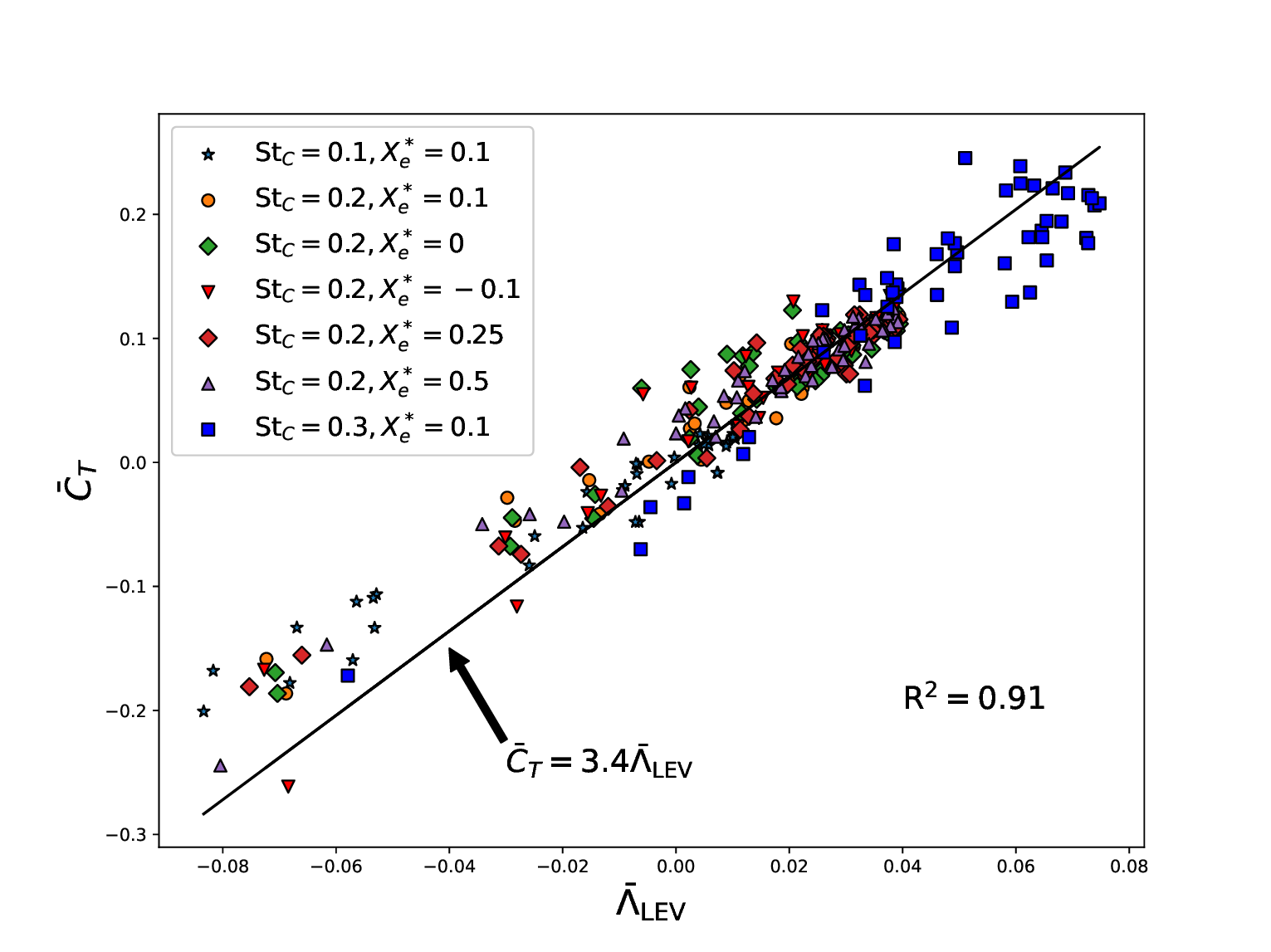}
    \caption{Plot showing the thrust coefficient ($\bar{C}_T$) and thrust factor ($\bar{\kappa}_\textrm{LEV}$) from each of 462 simulations and a linear fit through the data points. St$_C$ is the Strouhal number based on the foil chord and $X_e^*$ is the normalized location of the pivot on the foil. This plot is adapted from  Ref. \cite{raut2024hydrodynamic}}
    \label{kappa_psi}
\end{figure}

\clearpage

\end{document}